\documentclass[conference]{IEEEtran}
\IEEEoverridecommandlockouts
\usepackage{cite}
\usepackage{amsmath,amssymb,amsfonts}
\usepackage{algorithmic}
\usepackage{graphicx}
\usepackage{textcomp}
\usepackage[dvipsnames]{xcolor} 
\def\BibTeX{{\rm B\kern-.05em{\sc i\kern-.025em b}\kern-.08em
    T\kern-.1667em\lower.7ex\hbox{E}\kern-.125emX}}

\newcommand{\Section}{Section}
\usepackage{multirow}
\usepackage{algorithm}
\usepackage{amsmath}
\usepackage{subfigure}
\usepackage{mathtools}
\usepackage{bbm}
\usepackage{url}
\usepackage{booktabs}
\usepackage{color}
\usepackage[normalem]{ulem}

\begin{document}

\title{EdgePruner: Poisoned Edge Pruning in Graph Contrastive Learning}

\author{\IEEEauthorblockN{1\textsuperscript{st} Hiroya Kato}
  \IEEEauthorblockA{\textit{KDDI Research, Inc.} \\
    Japan\\
  }
  \\
  \IEEEauthorblockN{3\textsuperscript{rd} Seira Hidano}
  \IEEEauthorblockA{\textit{KDDI Research, Inc.} \\
    Japan\\
  }
  \and
  \IEEEauthorblockN{2\textsuperscript{nd} Kento Hasegawa}
  \IEEEauthorblockA{\textit{KDDI Research, Inc.} \\
    Japan\\
}
  \\
  \IEEEauthorblockN{4\textsuperscript{th} Kazuhide Fukushima}
  \IEEEauthorblockA{\textit{KDDI Research, Inc.} \\
    Japan\\
}
}

\maketitle

\newcommand{\ti}{adversarial edge pruning for GCL}
\newcommand{\prop}{EdgePruner}
\newcommand{\propnfs}{EdgePruner$^{\mathrm{NF}}$}
\newcommand{\baselinenfs}{Baseline$^{\mathrm{NF}}$}
\newcommand{\baseline}{Baseline}
\newcommand{\editing}[1]{\textcolor{red}{#1}}
\newcommand{\propboth}{EdgeModifier}
\newcommand{\propbothnf}{{EdgeModifier$^{\mathrm{NF}}$}}
\newcommand{\propbothbs}{EdgeModifier$^{\mathrm{b}}$}
\newcommand{\propbothnfbs}{EdgeModifier$^{\mathrm{b, NF}}$}

\newcommand{\cossim}[4]{\beta(\textit{h}^{#1}_{\mathit{#2}}, \textit{h}^{#3}_{\mathit{#4}})}
\newcommand{\Graph}{G}
\newcommand{\Node}{V}
\newcommand{\Edge}{E}
\newcommand{\AdjMatrix}{\textbf{\textit{A}}}
\newcommand{\NodeFeature}{\textbf{\textit{X}}}
\newcommand{\Encoder}{f}
\newcommand{\EmbeddingOne}{\textit{h}^{1}_{i}}
\newcommand{\EmbeddingTwo}{\textit{h}^{2}_{i}}
\newcommand{\View}[1]{G_#1=(\AdjMatrix_#1, \NodeFeature_#1)}
\newcommand{\Loss}{\mathcal{L}_{GCL}}
\newcommand{\PoisonedAdj}{\textbf{\textit{A}}^{'}}
\newcommand{\Parameters}{\theta}
\newcommand{\Augmentation}[1]{t_{#1}}
\newcommand{\Weight}{\textbf{\textit{W}}}
\newcommand{\Diagonal}{\textbf{\textit{D}}}
\newcommand{\Identity}{\textbf{\textit{I}}_{n}}
\newcommand{\EmbeddingMatrix}[1]{\textbf{\textit{H}}_{#1}}

\renewcommand{\algorithmicrequire}{\textbf{Input:}}
\renewcommand{\algorithmicensure}{\textbf{Output:}}
\newcommand{\PoisonedGraph}{G'}
\newcommand{\SanitizedAdj}{\textbf{\textit{S}}}
\newcommand{\UpdatedGraph}{G_{\mathrm{u}}}
\newcommand{\SanitizedGraph}{G_{s}}
\newcommand{\AugAdj}[1]{\SanitizedAdj^{k}_{#1}}
\newcommand{\AugGraph}[1]{G^{k}_{#1}}
\newcommand{\ViewK}[1]{(\SanitizedAdj^{k}_{#1}, \NodeFeature^{k}_{#1})}
\newcommand{\DoAugment}[1]{t^{k}_{#1}(\SanitizedAdj, \NodeFeature)}
\newcommand{\Gradient}[1]{\nabla^{k}_{#1}}
\newcommand{\GradientK}{\nabla^{k}}
\newcommand{\GradientSum}{\nabla_{\mathrm{total}}}
\newcommand{\SaniLoss}{\mathcal{L}_{s}}
\newcommand{\LossK}{\SaniLoss^{k}}
\newcommand{\CosSimilarity}{\mathrm{s}}
\newcommand{\Threshold}{T}
\newcommand{\InfoReg}{\mathcal{L}_{\mathrm{IR}}}
\newcommand{\MINLOSS}{\mathcal{L}_{\mathrm{min}}}
\newcommand{\OptimalSanitizedAdj}{\SanitizedAdj_{\mathrm{opt}}}
\newcommand{\Candidate}{\textbf{\textit{C}}}
\newcommand{\COMMENTS}[1]{\hfill $\triangleright$ #1}

\begin{abstract}
  Graph Contrastive Learning (GCL) is unsupervised graph representation learning that can obtain useful representation of unknown nodes.
  The node representation can be utilized as features of downstream tasks.
  However, GCL is vulnerable to poisoning attacks as with existing learning models.
  A state-of-the-art defense cannot sufficiently negate adverse effects by poisoned graphs although such a defense introduces adversarial training in the GCL.
  To achieve further improvement, pruning adversarial edges is important.
  To the best of our knowledge, the feasibility remains unexplored in the GCL domain.
  In this paper, we propose a simple defense for GCL, \emph{\prop}.
  We focus on the fact that the state-of-the-art poisoning attack on GCL tends to mainly add adversarial edges to create poisoned graphs, which means that pruning edges is important to sanitize the graphs.
  Thus, \prop\ prunes edges that contribute to minimizing the contrastive loss based on the node representation obtained after training on poisoned graphs by GCL.
  Furthermore, we focus on the fact that nodes with distinct features are connected by adversarial edges in poisoned graphs.
  Thus, we introduce feature similarity between neighboring nodes to help more appropriately determine adversarial edges.
  This similarity is helpful in further eliminating adverse effects from poisoned graphs on various datasets.
  Finally, \prop\ outputs a graph that yields the minimum contrastive loss as the sanitized graph.
  Our results demonstrate that pruning adversarial edges is feasible on six datasets.
   \prop\ can improve the accuracy of node classification under the attack by up to 5.55\% compared with that of the state-of-the-art defense.
   Moreover, we show that \prop\ is immune to an adaptive attack.
\end{abstract}


\section{Introduction}
The graph representation learning plays an important role in utilizing information from graph structured data such as social networks \cite{fan2019graph,zhang2021we} and e-commercial networks \cite{shchur2018pitfalls, zeng2019graphsaint} for the various purposes including community detection \cite{tu2018unified,chen2018supervised} and recommendation systems \cite{wu2019dual,wang2019neural,wang2020disentangled,wu2022graph}.
However, most existing graph representation learning methods are executed in a supervised or semi-supervised manner, which requires plentiful labeled data for training.
Unfortunately, the node labels of graph structured data are insufficient in quantity \cite{velickovic2019deep}.
This is because manually labeling nodes in large graphs is rarely realistic in practical situations.
Thus, unsupervised learning methods for graph structured data are needed.
In an early stage of research, traditional unsupervised methods such as DeepWalk \cite{perozzi2014deepwalk} and node2vec \cite{grover2016node2vec} yielded relatively insufficient performance compared with that of supervised methods.
However, recently, the classical information maximization (InfoMax) principle \cite{linsker1988self} has attracted renewed attention, and the contrastive learning has achieved great success in many fields such as computer vision \cite{gidaris2018unsupervised,bachman2019learning,he2020momentum,tian2020contrastive,li2020prototypical,caron2020unsupervised,chen2020simple} and natural language processing \cite{mnih2013learning,oord2018representation,kong2019mutual,fang2020cert,giorgi2020declutr,chi2021infoxlm}.
In particular, the breakthroughs of the contrastive learning in computer vision have motivated researchers to apply the similar techniques from visual representation learning to graph representation learning \cite{feng2022adversarial}.
To learn node representation in a graph without labels, several contrastive learning methods for graphs have been proposed.
In particular, such methods are called \emph{graph contrastive learning (GCL)} in this paper.
Recent GCL achieves comparable performance with supervised methods by introducing various techniques, including different view generation via stochastic augmentations and a well-considered contrastive loss.
The main goal of GCL is to learn an encoder that maps nodes in a graph into low-dimensional numerical vectors (the vectors are called embeddings).
Useful embeddings can be acquired by training the encoder so that only similar nodes in the input space are close in the embedding space. 
The embeddings can be utilized as features for many downstream tasks such as node classification.
Thus, GCL is helpful in obtaining useful representation of unseen nodes.

However, a recent study \cite{zhang2022unsupervised} shows that the GCL models are vulnerable to poisoning attacks.
To make matters worse, the state-of-the-art attack called CLGA \cite{zhang2022unsupervised} can degrade the performance of many downstream tasks by contaminating embeddings output from an encoder trained by GCL.
Thus, CLGA may cause a severe obstacle that prevents GCL from being widely utilized.
As a state-of-the-art defensive method in the GCL domain, an effective GCL method called ARIEL \cite{feng2022adversarial} has been proposed recently.
In ARIEL, another view called the adversarial view is introduced so as to make the GCL model robust to poisoning attacks.
As a result, ARIEL can improve the quality of the embeddings based on poisoned graphs.
However, ARIEL only tries to learn a robust model when a poisoned graph is given.
Accordingly, ARIEL achieves suboptimal performance on poisoned graphs, which means that there is still room for improvement to the quality of the embeddings.
To achieve further improvement, we argue that sanitizing the poisoned graph is needed.
Sanitizing means pruning adversarial edges in a poisoned graph in this paper.
To the best of our knowledge, there has been no study that explores the feasibility of sanitizing poisoned graphs in the GCL domain.

In this paper, we propose a simple defense called \emph{\prop}.
We focus on the fact that the state-of-the-art attack on GCL, namely CLGA tends to mainly add adversarial edges to create poisoned graphs.
On this basis, we consider that the pruning edges is important to sanitize poisoned graphs. 
\prop\ prunes edges that contribute to minimizing the contrastive loss on the basis of gradients of the loss.
By doing this, sanitizing poisoned graphs can be expected.
Furthermore, to help determine more appropriate edges to prune, we focus on the fact that adversarial edges tend to connect nodes for which the features are dissimilar in poisoned graphs, which is shown in other work \cite{wu2019adversarial,jin2020graph}.
This is because connecting such nodes is effective in confusing the relationship between neighboring nodes in a graph from the perspective of attacks.
Thus, in light of this tendency, we introduce feature similarity between neighboring nodes as additional information to determine adversarial edges connecting nodes whose features are distinct.
This similarity is helpful in further eliminating adverse effects from poisoned graphs on various datasets, which further improves the quality of the embeddings.
Finally, \prop\ outputs a graph that yields the minimum contrastive loss as the sanitized graph.
The sanitized graphs are fed into GCL methods for training the encoder.

\smallskip
\noindent \textbf{Our contributions.} The main contributions of this work are as follows:
\begin{enumerate}
  \item We propose a simple and effective pruning method against poisoning attacks for the GCL.
    Our pruning method sanitizes poisoned graphs by pruning edges that contribute to minimizing the contrastive loss.
    We formulate our pruning method as the optimization problem.
  \item We introduce feature similarity between neighboring nodes to prune adversarial edges connecting nodes with distinct features.
    We also formulate our pruning method with the feature similarity as the optimization problem.
    We show that the similarity is helpful in further eliminating adverse effects from poisoned graphs on various datasets.
  \item We conduct extensive experiments to demonstrate the effects of \prop\ for clean graphs and poisoned graphs created by the CLGA and an adaptive attack.
  \item Our experimental results demonstrate that \prop\ can eliminate the detrimental effects from the poisoned graphs on six datasets while maintaining acceptable accuracies on clean graphs. 
  In particulr, \prop\ can improve accuracy by up to 9.60\% compared with that of a GCL method without defense.
    Moreover, we show that EdgePruner is immune to an adaptive attack.
    To the best of our knowledge, this work first shows the feasibility of pruning adversarial edges in poisoned graphs in the GCL domain.
\end{enumerate}

\section{Related Work} \label{sec:related}
Most existing graph representation learning methods are supervised or semi-supervised ones \cite{scarselli2008graph,velivckovic2017graph,hamilton2017inductive,kipf2016semi,xu2018powerful}.
However, the node labels of large graphs in the real world are difficult to obtain \cite{velickovic2019deep}.
This is because manually labeling nodes in such graphs is rarely realistic in practical situations.
Therefore, in the graph domain, several GCL methods, which can learn graph representation without labels have been proposed. 
The main goal of GCL is to learn an encoder that converts nodes in a graph into low-dimensional embeddings without labels.
For example, as an encoder, a two-layer (graph convolutional network) GCN \cite{kipf2016semi} is utilized.
An encoder is trained so that only similar nodes in the input space are close in the embedding space.
The embeddings produced by the trained encoder can be utilized as features for downstream tasks such as node classification and link prediction.

\subsection{Graph Contrastive Learning}
To learn node representation in a graph without labels, several GCL models have been proposed \cite{velivckovic2018deep, hassani2020contrastive, you2020graph, qiu2020gcc, zhu2020deep, zhu2021graph}.
Velivckovic et al. \cite{velivckovic2018deep} propose deep graph InfoMax (DGI), which is a general approach for unsupervised graph learning based on mutual information, rather than random walks.
DGI maximizes mutual information between patch representations and corresponding high-level summaries of graphs derived by using graph convolutional network architectures.
The patch representations are vectors that express local information about the graph centered around a node rather than just the node.
According to the results in \cite{velivckovic2018deep}, DGI yields higher performance on downstream tasks compared with that of a traditional unsupervised method, DeepWalk \cite{perozzi2014deepwalk}.

To supplement the input graph with more global information, Hassani and Khasahmadi propose contrastive multi-view graph representation learning (called MVGRL in this paper) \cite{hassani2020contrastive}.
In MVGRL, graph diffusion is introduced into the GCL approach, and graph views are obtained by uniformly sampling subgraphs.
Then, MVGRL contrasts node representations to global embeddings across the two views.
As a result, MVGRL outperforms DGI. 
You et al. propose GraphCL \cite{you2020graph} and design four types of graph augmentations, namely, node dropping, edge perturbation, attribute masking, and sampling subgraphs.
The impact of various combinations of these graph augmentations on multiple datasets is systematically studied.
Qiu et al. propose the graph contrastive coding (GCC) framework to learn structural representations across graphs \cite{qiu2020gcc}.
GCC aims to distinguish between subgraphs sampled from a certain node and subgraphs sampled from other nodes.
Since GCC does not assume that nodes and subgraphs come from the same graph, the graph encoder is designed to capture universal patterns across different input graphs. 
In other words, the pretrained GCC model can be applied to unseen graphs for downstream tasks such as node classification.
Zhu et al. propose GRACE \cite{zhu2020deep}, which generates two correlated graph views by randomly performing corruption.
The model is trained by using a contrastive loss to maximize the agreement between node embeddings in these two views.
They also consider random augmentation at both topological and node attribute levels so as to accelerate optimization of the contrastive loss.
The augmentation includes removing edges and masking features to provide diverse contexts for nodes in different views.
Zhu et al. propose graph contrastive learning with adaptive augmentation (GCA) \cite{zhu2021graph}.
GCA introduces adaptive augmentation that perturbs both node features and edges in accordance with their importance.
They argue that augmentation schemes used in the existing methods suffer from suboptimal performance because most existing methods adopt uniform data augmentation schemes, such as uniformly dropping edges and uniformly shuffling features.
On the topological level, new augmentation schemes based on node centrality measures are designed to identify important edges.
Furthermore, noise is added to node attributes by randomly masking some node features with zeros.
As a result, GCA further improves GRACE.

\subsection{Poisoning Attack on GCL} \label{sec:poisoning_attack}
Although GCL is promising for learning useful representation of unseen nodes, a recent study \cite{zhang2022unsupervised} shows that the GCL models are also vulnerable to poisoning attacks as with existing learning methods.
Almost all of the existing attacks are mainly intended for supervised learning methods such as GCN.
In other words, they are supervised attacks and require labels to create poisoned graphs.
PGD and MinMax \cite{xu2019topology} optimize the negative cross-entropy loss on the basis of gradients.
Nettack \cite{zugner2018adversarial} iteratively selects edges to alter by calculating the score of each possible alteration.
Mettack \cite{zugner_adversarial_2019} uses the gradient of the classification loss with respect to the adjacency matrix to select the edges to change.

However, in real-world scenarios, it is difficult to attack GCL with the abovementioned supervised attacks because the labels of large graphs are difficult to acquire.
This motivated researchers to demonstrate whether unsupervised attacks\footnote{Unsupervised attacks mean that attackers succeed in creating poisoned graphs without labels of nodes} targeting at GCL are feasible without labels or not.
Bojchevisk and Stephan propose a graph poisoning attack based on random walks~\cite{bojchevski2019adversarial}.
They devise efficient adversarial perturbations that poison the network structure of graphs.
Their poisoning attack has a negative effect on both the quality of the node embeddings and the downstream tasks.
However, their attack cannot work well for GCA that is the latest GCL model because that attack relies only on random walks.
Zhang et al. \cite{zhang2022unsupervised} propose an unsupervised attack based on gradient ascent on the GCL for node embeddings, which is called CLGA.
CLGA is intended for GCL and computes the gradient of the contrastive loss w.r.t. the adjacency matrix.
After that, the edges with the largest gradients are flipped (deletion or addition).
According to the experiments in \cite{zhang2022unsupervised}, CLGA outperforms the existing unsupervised attack \cite{bojchevski2019adversarial}.
In addition to that, CLGA has comparable performance with some of the existing supervised attacks.
These results mean that the emergent poisoning attack can prevent GCL from being widely utilized from now on.
Therefore, developing defense and robust GCL models that resist poisoning attacks is of significant importance.

\subsection{Defenses against Poisoning Attack on GCL} \label{sec:defense}
Recently, Feng et al. have proposed a GCL method called ARIEL \cite{feng2022adversarial} that is robust to poisoning attacks.
In ARIEL, an adversarial view created by an existing attack \cite{madry2017towards} is introduced so as to make the GCL model robust to poisoning attacks.
In other words, the technique that is equivalent to adversarial training is utilized in ARIEL.
The adversarial view is treated as another view to assimilate adversarial training to the GCL.
The adversarial contrastive loss is newly defined as the contrastive loss between one of the two views and the adversarial view.
Although adversarial training in ARIEL may effectively improve the robustness of the model to poisoned graphs, the progress in such hard training may be stagnant in bad parameter area at an early stage.
To realize stable training, ARIEL introduces one additional constraint called information regularization.
Furthermore, ARIEL adopts two additional techniques, namely subgraph sampling and curriculum learning.
The subgraph sampling can reduce the computational cost because the gradient derivation on the entire graph is avoided.
On the other hand, the curriculum learning contributes to making GCL become harder gradually by increasing the portion of the adversarial contrastive loss as the training progresses.
Accordingly, ARIEL outperforms the existing GCL methods in the node classification task on poisoned graphs, which is further improvement in the robustness of the GCL.
Strictly speaking, ARIEL is not against a GCL poisoning attack. 
However, compared with GCA, ARIEL can counter the GCL poisoning attack, namely CLGA to some extent according to our experiments. 
Furthermore, ARIEL can mitigate bad effects of poisoned graphs without utilizing labels. 
This corresponds to our assumed situation that is defending GCL from poisoning attacks without depending on labels, which is described in \Section~\ref{sec:problem_motivation}.
Thus, we regard ARIEL as a state-of-the-art defense in the GCL domain.
ARIEL is a promising defense against CLGA.
However, ARIEL only tries to learn a robust model by introducing the adversarial view when a poisoned graph is given.
Thus, as shown in \Section~\ref{sec:experiment}, ARIEL cannot sufficiently eliminate adverse effects from poisoned graphs, which means that there is still room for improvement to the quality of the embeddings.

On the other hand, \prop\ can further eliminate adverse effects from poisoned graphs.
Our method focuses on eliminating adverse effects by pruning adversarial edges rather than learning robust parameters of the encoder by the GCL, which is the different point from ARIEL.

\section{Preliminaries} \label{sec:preliminary}
\subsection{GCL for Node Embeddings}

In this work, we focus on node-level GCL.
Let $\Graph =(\AdjMatrix, \NodeFeature)$ denote a graph, where $\AdjMatrix \in \mathbb{R}^{N \times N}$ is the adjacency matrix, and $\NodeFeature \in \mathbb{R}^{N \times d}$ is the node feature matrix.
$N$ is the number of nodes, and $d$ is the number of features.
The objective of GCL is to learn an encoder $\Encoder(\AdjMatrix, \NodeFeature)$ that outputs node embeddings.
In many cases \cite{zhu2021graph,zhang2022unsupervised,feng2022adversarial}, GCN \cite{kipf2016semi} is utilized as the encoder $\Encoder$ in GCL due to its simplicity and the stable performance.
In this work, we also employ a two-layer GCN as with the other work.
The overview of GCN is explained in Appendix~\ref{appendix:gcn}.
As for GCL method, we utilize GCA \cite{zhu2021graph}, which is the state-of-the-art GCL in our experiments.
The details of GCA are described in Appendix~\ref{appendix:gca}.

\subsection{Unsupervised poisoning attack on GCL}
We focus on the state-of-the-art attack in the GCL domain, namely CLGA in this work.
CLGA is the untargeted poisoning attack that deteriorates the overall classification performance for samples in every class.
The purpose of CLGA is to poison graphs so that the quality of the embeddings learned by GCL is degraded.
As a result, the poisoned embeddings cause worse performance in downstream tasks.
In what follows, we explain the overview of CLGA.

\smallskip
\noindent \textbf{Overview of CLGA.}
CLGA only modifies edges and does not change node features.
In other words, CLGA creates only the poisoned adjacency matrix $\PoisonedAdj$.
CLGA creates $\PoisonedAdj$ in the iterative manner so as to locate more informative edges.
Once an edge is modified at each iteration, the modified adjacency matrix is used to retrain the encoder to obtain gradients in the next iteration.
In an attempt to create $\PoisonedAdj$, CLGA tries to solve the following optimization problem 
\begin{equation}
\begin{gathered}
    \underset{\PoisonedAdj}{\mathrm{max}}\ \mathcal{L}(\Encoder_{\Parameters'} (\AdjMatrix_{1}, \NodeFeature_{1}), \Encoder_{\Parameters'} (\AdjMatrix_{2}, \NodeFeature_{2})),  
  \\
    \mathrm{s.t.}\ \Parameters' = \underset{\Parameters}{\mathrm{argmin}}\ \mathcal{L}(\Encoder_{\Parameters} (\AdjMatrix_{1}, \NodeFeature_{1}), \Encoder_{\Parameters} (\AdjMatrix_{2}, \NodeFeature_{2})), 
  \\
  (\AdjMatrix_1, \NodeFeature_1) = \Augmentation{1}(\PoisonedAdj, \NodeFeature), 
  (\AdjMatrix_2, \NodeFeature_2) = \Augmentation{2}(\PoisonedAdj, \NodeFeature), \| \AdjMatrix - \PoisonedAdj \|_{F}^{2} = \sigma.
\end{gathered}
\end{equation}
The Frobenius norm of $\AdjMatrix$ is defined by $\|\AdjMatrix\|_{F} = \sqrt{\sum_{i,j}{\AdjMatrix[i,j]}}$.
In addition to that, $\Augmentation{1}$ and  $\Augmentation{2}$ are two stochastic augmentation procedures.
When it is assumed that GCA is attacked, $\mathcal{L}$ is equivalent to the contrastive loss in Eq.~(\ref{eq:loss}) in Appendix~\ref{appendix:gca}.
$\Parameters$ is parameters of $\Encoder$.
The number of pruned edges is bounded by a given threshold $\sigma$.

\subsection{Threat Model} 
\begin{figure}
  \includegraphics[scale=0.30]{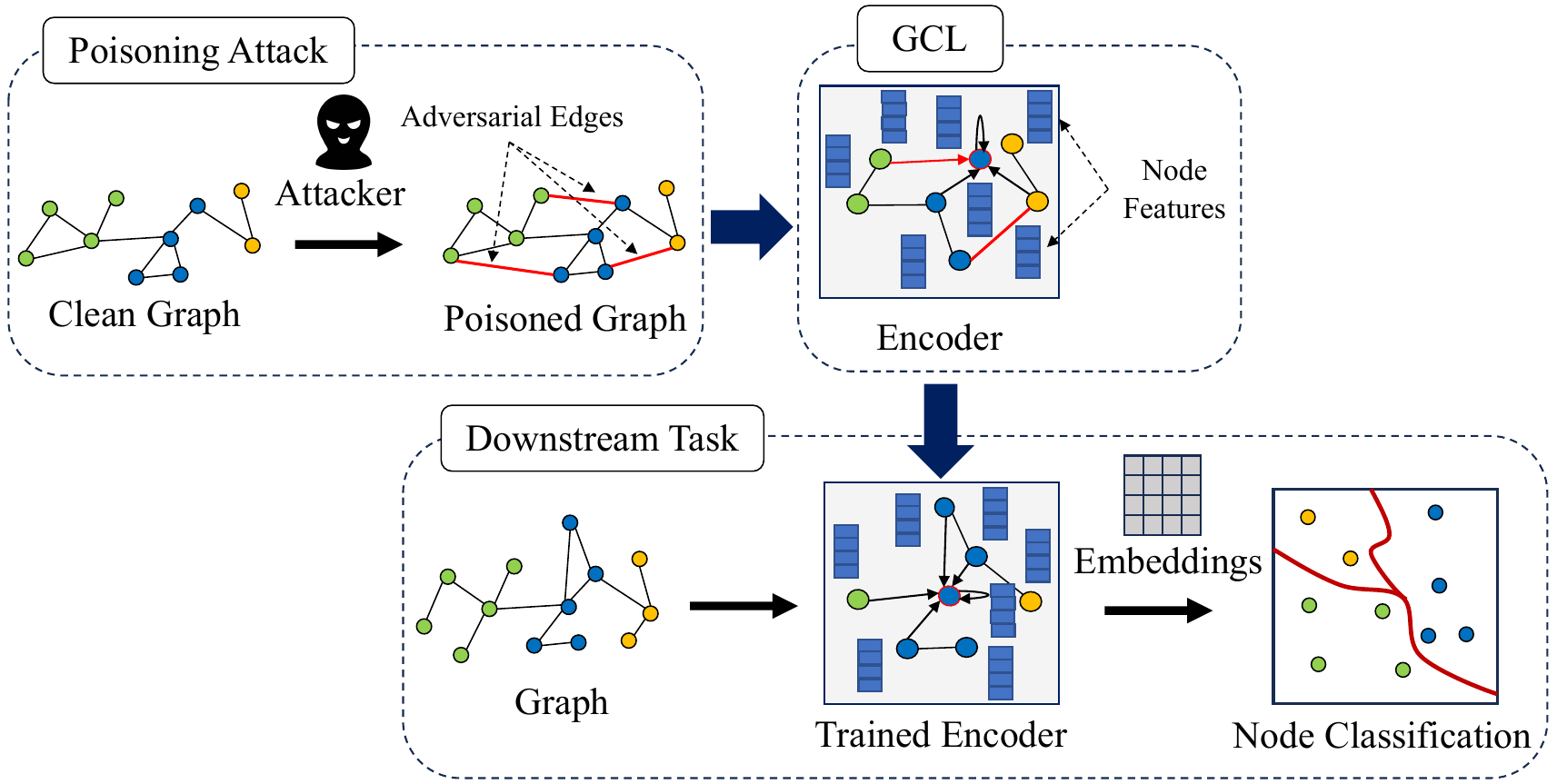}
  \caption{Threat model.}
  \label{fig:threat_model}
\end{figure}
\figurename~\ref{fig:threat_model} shows the threat model in this work.
The encoder is trained by GCL on unlabeled graphs in order to facilitate utilizing them in downstream tasks such as node classification.
The encoder may be trained on graphs published by a third party.
Downstream tasks utilize the embeddings obtained through inputting nodes in a graph into the trained encoder as features.

\smallskip
\noindent \textbf{Attacker's goal.}
Attacker's goal is to extensively launch attack on many downstream tasks.
Considering the efficiency of attacks, it is desirable for the attacker to directly contaminate the encoder trained GCL rather than downstream tasks.
This is why the attacker conducts the poisoning attack on GCL.

\smallskip
\noindent \textbf{Attacker's capability.}
The attacker has no access to both the classification models and graph datasets utilized in the downstream tasks.
Meanwhile, the attacker knows the model parameters of target GCL for some reasons such as information leakage, which means the worst case for defenders.
Thus, the attacker can reproduce the target GCL model.
To create the poisoned graph, the attacker modifies only edges in a graph so that the loss of the reproduced model is maximized by CLGA.
The poisoned graphs are published by the attacker so as to contaminate encoders.
If the attacker successfully poisons the target encoder, the resulting embeddings are degraded.
Accordingly, the attacker can indirectly have detrimental effects on myriads of downstream tasks, which means the attack success.

\section{Problem Statement and Motivation} \label{sec:problem_motivation}
As mentioned in \Section~\ref{sec:poisoning_attack}, we consider that the poisoning attack on GCL becomes a severe obstacle.
However, the studies regarding defensive methods for GCL are not adequately explored.
To defend GCL, it is preferable that defenders counter the attack without depending on labels as much as possible.
Thus, in this work, we assume the situation where defenders cannot utilize abundant labels of graphs.
Although ARIEL can mitigate the bad effect of poisoned graphs, there is still room for improvement.
To achieve further improvement, we argue that sanitizing the poisoned graph is needed.
Sanitizing means pruning adversarial edges from graphs in this work.
Furthermore, we assume that edge pruning is conducted just before training an encoder rather than training classification models in downstream tasks.
This is because defending training encoders is more efficient in light of myriads of downstream tasks.
To the best of our knowledge, there has been no study that focuses on adversarial edge pruning in the GCL domain.
Therefore, since the feasibility still remains largely unexplored, we decided to work on that.

\section{Proposed Method} \label{sec:proposal}
\begin{figure*}
  \centering
  \includegraphics[scale=0.38]{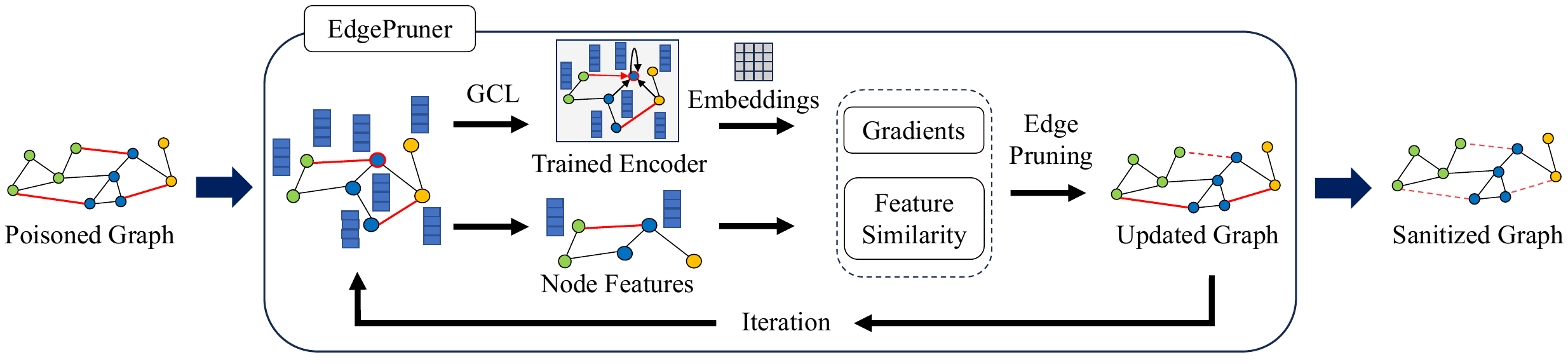}
  \caption{Overview of \prop.}
  \label{fig:proposal_overview}
\end{figure*}

\subsection{Overview} \label{subsec:overview}
In this paper, we propose \prop, which is  \ti.
We focus on the fact that the poisoning attack on GCL tends to mainly add adversarial edges to create poisoned graphs.
On the basis of this fact, we consider that the deletion of edges is important to sanitize poisoned graphs.
One promising way for determining which edges should be deleted is to utilize gradients of the contrastive loss based on embeddings from the encoder that trains on a poisoned graph.
By pruning edges that contribute to minimizing the contrastive loss, sanitizing poisoned graphs can be expected.
However, there are cases where it is difficult to decide edges to prune because the trained encoder is influenced by adversarial edges in the poisoned graph.
As a result, the contrastive loss based on the embeddings from the encoder may not be totally reliable.
Furthermore, since the node labels are not available in GCL, counting only on the encoder trained on poisoned graphs is not necessarily desirable.
Thus, additional information that is not influenced by adversarial edges is needed.
In order to help determine the appropriate edges, we focus on the fact that adversarial edges tend to connect nodes for which the features are dissimilar.
This is because connecting such nodes is effective in confusing the relationship between neighboring nodes in a graph.
As a result, embeddings are poisoned, which results in degrading the performance of the downstream tasks.
In fact, recently, it has been demonstrated that adversarial attacks on graphs tend to connect nodes with distinct features \cite{wu2019adversarial,jin2020graph}.
Thus, in light of this tendency, we utilize feature similarity between neighboring nodes connected to each other in order to prune adversarial edges connecting nodes with distinct features.
This similarity is helpful in further eliminating adverse effects from poisoned graphs.
Finally, EdgePruner outputs a graph that yields the minimum contrastive loss as the sanitized graph. 
\figurename~\ref{fig:proposal_overview} shows the overview of \prop.

In what follows, we first validate how edges are changed when poisoned graphs are created and formulate an optimization problem that \prop\ solves.
After that, we elaborate on the usefulness of feature similarity between neighboring nodes.
Finally, the algorithm of \prop\ is explained.

\subsection{Edge Modification on Poisoned Graphs}
To validate the tendency that adversarial edges are principally added for creating poisoned graphs, we inspect the number of edges that are modified when poisoned graphs are created.
\tablename~\ref{tab:adversarial_edges} shows the number of modified edges when poisoned graphs are created by CLGA.
\begin{table*}[t]
  \caption{The number of edges that are added and deleted when poisoned graphs are created by CLGA.}
  \label{tab:adversarial_edges}
  \centering
  \begin{tabular}{c|c|c|c|c|c|c|c|c|c|c|c|c|c|c|c|c|c|c}
    \toprule
     & \multicolumn{3}{c|}{Cora} & \multicolumn{3}{c|}{CiteSeer} & \multicolumn{3}{c|}{Amazon-Computers} & \multicolumn{3}{c|}{Amazon-Photo}& \multicolumn{3}{c|}{Coauthor-CS} & \multicolumn{3}{c}{Coauthor-Physics} \\
    \cmidrule{2-19}
    & 1\% & 5\% & 10\% & 1\% & 5\% & 10\% & 1\% & 5\% & 10\% & 1\% & 5\% & 10\% & 1\% & 5\% & 10\% & 1\% & 5\% & 10\% \\
    \midrule
    Added edges & 52 & 263 & 528 & 45 & 227 & 445 & 355 & 1698 & 3235 & 496 & 2477 & 4927 & 61 & 309 & 618 & 52 & 260 & 523 \\
    Deleted edges & 0 & 0 & 0 & 0 & 0 & 0 & 1 & 85 & 333 & 0 & 3 & 33 & 0 & 0 & 1 & 0 & 2 & 2 \\
  \bottomrule
\end{tabular}
\end{table*}
In this inspection, we utilize six datasets, namely Cora, CiteSeer \cite{yang2016revisiting}, Amazon-Computers, Amazon-Photo \cite{shchur2018pitfalls}, Coauthor-CS, and Coauthor-Physics \cite{shchur2018pitfalls}.
Note that we utilize poisoned graphs created from subgraphs of 5,000 nodes on Amazon-Computers, Amazon-Photo, Coauthor-CS, and Coauthor-Physics as with the experimental setup in \cite{feng2022adversarial} because we could not create the poisoned graphs based on their entire graphs due to memory limitations.
Detailed information about these datasets is described in \Section~\ref{sec:experiment}.
The poisoned graphs in \tablename~\ref{tab:adversarial_edges} are the ones created by maximizing the loss of GCA.
As for ARIEL, we observe a similar result, which is shown in Appendix~\ref{appendix:num_edges_ariel}.
As shown in \tablename~\ref{tab:adversarial_edges}, it is obvious that poisoned graphs are created by adding edges to clean graphs in most cases.
In particular, as for Cora and CiteSeer, poisoned graphs are created only by adding edges.
Therefore, we consider that the deletion of edges is more important than the addition so as to sanitize poisoned graphs by unsupervised attacks on GCL. 

\prop\ deletes edges on the basis of gradient descent on the adjacency matrix.
In other words, \prop\ converts $\PoisonedAdj$ to the sanitized adjacency matrix $\SanitizedAdj$ through edge pruning so that a contrastive loss $\SaniLoss$ of an encoder $\Encoder$ that we defend is minimized.
Thus, \prop\ solves the following optimization problem formulated as 
\begin{equation} \label{eq:opt_problem1}
\begin{gathered}
  \underset{\SanitizedAdj}{\mathrm{min}}\ \SaniLoss(\Parameters', \SanitizedAdj, \NodeFeature),  \\
  \mathrm{s.t.}\ \Parameters' = \underset{\Parameters}{\mathrm{argmin}}\ \SaniLoss(\Parameters, \SanitizedAdj, \NodeFeature),\ \|\PoisonedAdj - \SanitizedAdj\|_{F}^{2} \leq \sigma^{'},
\end{gathered}
\end{equation}
where $\SaniLoss$ is defined as
\begin{equation} \label{eq:sanitized_loss}
\begin{split}
            \SaniLoss(\Parameters', \SanitizedAdj, \NodeFeature) =\ & \Loss(\EmbeddingMatrix{\Parameters^{'}}^{1}, \EmbeddingMatrix{\Parameters^{'}}^{2}) \\
                                                                    & + \epsilon_{1} \cdot \Loss(\EmbeddingMatrix{\Parameters^{'}}^{1}, \EmbeddingMatrix{\Parameters^{'}}^{3}) + \epsilon_{2} \cdot \InfoReg.
\end{split}
\end{equation}
Note that $\EmbeddingMatrix{\Parameters^{'}}^{m}$ is defined in Appendix~\ref{appendix:gca}.
$\InfoReg$ is information regularization introduced in ARIEL.
Also, $\epsilon_1$ and $\epsilon_2$ are predefined parameters to control adversarial contrastive loss and information regularization used in ARIEL.
When $\epsilon_1=0$ and $\epsilon_2=0$, $\SaniLoss$ is equivalent to the contrastive loss of GCA.
The number of modified edges is bounded by a threshold $\sigma^{'}$.

In order to select informative edges, we execute $K$ times calculations of $\SaniLoss$.
On the basis of $\SanitizedAdj$ and $\NodeFeature$, multiple views $\AugGraph{m} = \ViewK{m}$ are obtained via the stochastic augmentations, which are repeated $K$ times.
The value of $m$ is an index for identifying $M$ views generated in GCL methods.
For example, since GCA generates two views, $M=2$.
On the other hand, since ARIEL requires the adversarial view in addition to the two views, $M=3$.
At the $k$-th augmentation, the contrastive loss $\LossK$ based on $\AugGraph{m}$ is computed.
With respect to $\AugAdj{m}$, gradient matrices of $\LossK$, namely $\Gradient{m} \in \mathbb{R}^{N \times N}$ are calculated.
When $\Encoder(\SanitizedAdj, \NodeFeature)$ is a differentiable encoder such as GCN, we can calculate $\Gradient{m}$ as 
\begin{equation}
  \Gradient{m} = \frac{\partial \LossK}{\partial \AugAdj{m}} = \frac{\partial \LossK}{\partial \Encoder \ViewK{m}} \cdot \frac{\partial \Encoder \ViewK{m}}{\partial \AugAdj{m}}.
\end{equation}

To alleviate the bias caused by the stochastic augmentation, we adopt two techniques that are introduced in CLGA \cite{zhang2022unsupervised}. 
The first one is adding up $\Gradient{m}$.
The gradient matrix at the $k$-th augmentation is regarded as
\begin{equation}
  \nabla^{k} = \sum_{m}^{M}{\Gradient{m}},
\end{equation}
where $M$ is the number of augmented views used in an assumed GCL.
The other is adding up $\GradientK$ over $K$ times iteration, which means $\GradientSum = \sum_{k=1}^{K}{\GradientK}$.
This total gradient matrix $\GradientSum$ is utilized to select pruned edges.

\prop\ prunes edges that meet a condition regarding the presence of edges and the correct direction of gradients.
Pruning an edge means changing a value from 1 to 0 in the adjacency matrix.
Thus, the condition is represented as 
\begin{equation}
  \label{eq:del_condition}
  \SanitizedAdj[i,j] = 1 \land  \GradientSum[i, j] > 0.
\end{equation}
If an existent edge (which means the corresponding value is 1 in the adjacency matrix) has a positive gradient, changing the value from 1 to 0 would decrease the loss, which means deleting an edge.
Our edge pruning is mainly conducted by deleting an edge out of edges that meet Eq.~(\ref{eq:del_condition}) one by one.
After an edge is deleted, $\SanitizedAdj$ is checked to see whether the loss $\SaniLoss$ is minimum at that time.
Let $\MINLOSS$ denote the minimum loss during our pruning procedures.
If $\SaniLoss$ is less than $\MINLOSS$, $\SanitizedAdj$ is saved as the optimal adjacency matrix $\OptimalSanitizedAdj$, which should be returned as output.
The updated graph $\UpdatedGraph=(\SanitizedAdj, \NodeFeature)$ is fed into $\Encoder$ to retrain.
Once the encoder is retrained, $\GradientSum$ based on node embeddings from the trained encoder is recalculated.
Then, another edge is pruned in the next iteration.
By iterating the above procedures, the poisoned graph is gradually sanitized.
Finally, the sanitized graph $\SanitizedGraph = (\OptimalSanitizedAdj, \NodeFeature)$ is outputted.

\subsection{Features Similarity of Neighboring nodes}
Pruning adversarial edges is not always an easy task if we rely only on gradients calculated on poisoned graphs.
This is because training the encoder is influenced by adversarial edges in poisoned graphs at each iteration, which interrupts selecting edges to prune appropriately.
To help determine the appropriate edges, we focus on the feature similarity between neighboring nodes connected to each other.
As mentioned in \Section~\ref{subsec:overview}, nodes that have distinct features are connected when the poisoned graph is created.
We inspect the feature similarity of neighboring nodes connected by clean edges or adversarial ones.
\tablename~\ref{tab:sim_feature} shows the average cosine similarity of neighboring nodes in poisoned graphs based on the loss of GCA.
It is observed that the poisoned graphs based on the loss of ARIEL also have similar tendency in our preliminary experiment.
In \tablename~\ref{tab:sim_feature}, we also utilize the poisoned graphs created from subgraphs of Amazon-Computers, Amazon-Photo, Coauthor-CS, and Coauthor-Physics.
\begin{table*}[t]
  \caption{The average cosine similarity of nodes connected by clean edges or adversarial ones in the poisoned graphs with standard deviation.}
  \label{tab:sim_feature}
  \centering
  \begin{tabular}{c|c|c|c|c|c|c}
    \toprule
     & Cora & CiteSeer & Amazon-Computers & Amazon-Photo & Coauthor-CS & Coauthor-Physics \\
    \midrule
    Clean edges & 0.167 $\pm$ 0.127 & 0.191 $\pm$ 0.140 & 0.495 $\pm$ 0.209 & 0.489 $\pm$ 0.202 & 0.297 $\pm$ 0.133 & 0.352 $\pm$ 0.154 \\
    Adversarial edges & 0.029 $\pm$ 0.048 & 0.021 $\pm$ 0.029 & 0.297 $\pm$ 0.087 &  0.297 $\pm$ 0.098 &  0.011 $\pm$ 0.018 & 0.025 $\pm$ 0.032\\
  \bottomrule
\end{tabular}
\end{table*}
As shown in \tablename~\ref{tab:sim_feature}, the average cosine similarities of nodes connected by adversarial edges are small compared with those of nodes connected by clean edges.
Furthermore, distribution of the cosine similarity on the six datasets are shown in Appendix~\ref{appendix:cos_hist}.
These inspection results demonstrate that poisoned graphs tend to be created by connecting nodes whose features are distinct.
Hence, we introduce the cosine similarity of features of neighboring nodes as additional information for appropriately determining edges to prune.
If node features are available, \prop\ deletes an edge that meets Eq.~(\ref{eq:del_condition}) and connects two nodes whose features are distinct to some extent at each iteration. 
In other words, the condition regarding feature similarity is represented as
\begin{equation}
  \label{eq:fsim_condition}
\CosSimilarity(\NodeFeature[i,:], \NodeFeature[j,:]) < \Threshold,
\end{equation}
where $i$ and $j$ are indices of nodes.
Also, $\CosSimilarity(\cdot, \cdot)$ and $\Threshold$ are the cosine similarity and a threshold, respectively.
In practical situations, $\Threshold$ should be tuned depending on parameters of GCL methods and datasets.
Thus, when node features are available, instead of Eq.~(\ref{eq:opt_problem1}), \prop\ solves the following optimization problem
\begin{equation} \label{eq:opt_problem2}
\begin{gathered}
  \underset{\SanitizedAdj}{\mathrm{min}}\ \SaniLoss(\Parameters', \SanitizedAdj, \NodeFeature),\ \mathrm{s.t.}\ \Parameters' = \underset{\Parameters}{\mathrm{argmin}}\ \SaniLoss(\Parameters, \SanitizedAdj, \NodeFeature), \\
  \|\PoisonedAdj - \SanitizedAdj\|_{F}^{2} \leq \sigma^{'}, \overline{\CosSimilarity}(\SanitizedAdj) > \overline{\CosSimilarity}(\PoisonedAdj), 
\end{gathered}
\end{equation}
where $\overline{\CosSimilarity}$ is defined as 
\begin{equation}
\begin{lgathered}
  \overline{\CosSimilarity}(\AdjMatrix) = \frac{{\sum_{i,j}{\CosSimilarity(\NodeFeature[i,:], \NodeFeature[j,:])}} \cdot \AdjMatrix[i, j]}{\|\AdjMatrix\|_{F}^{2}}.
\end{lgathered}
\end{equation}

\subsection{Algorithm}
In this subsection, we concisely describe the algorithm of \prop.
Algorithm~\ref{alg1} shows the algorithm of \prop.
\begin{figure}[t]
\begin{algorithm}[H]
    \caption{\prop}
    \label{alg1}
    \begin{algorithmic}[1]    
    \REQUIRE Poisoned Graph $\PoisonedGraph = (\PoisonedAdj, \NodeFeature)$
    \STATE $n \leftarrow 0$
    \STATE Initialize the sanitized adjacency matrix $\SanitizedAdj \leftarrow \PoisonedAdj$
    \STATE Initialize the optimal matrix $\OptimalSanitizedAdj \leftarrow \SanitizedAdj$ 
    \STATE Initialize the minimum loss $\MINLOSS  \leftarrow \infty$
    \WHILE{$n < N$} 
    \STATE Train $\Encoder$ with $\SanitizedAdj$ and $\NodeFeature$ from scratch
    \STATE Initialize gradients $\GradientSum \leftarrow 0$
     \FOR {$k = 1$ to $K$} 
     \STATE Obtain $m$ views $\AugGraph{m} = \ViewK{m}$ 
     \STATE Compute $\LossK$ based on the trained $\Encoder$ via Eq.~(\ref{eq:sanitized_loss}) 
     \STATE $\Gradient{m} \leftarrow \frac{\partial \LossK}{\partial \AugAdj{m}}$  \COMMENTS{Compute the gradient.}
       \STATE $\nabla^{k} \leftarrow \sum_{m}^{M}{\Gradient{m}}$ 
       \STATE $\GradientSum \leftarrow \GradientSum + \nabla^{k}$  
       \ENDFOR
     \STATE Create a set of candidates $\Candidate$ based on Eq.~(\ref{eq:del_condition}) and Eq.~(\ref{eq:fsim_condition})
     \IF{$\Candidate \neq \varnothing$}
      \STATE Delete $\Candidate[i,j]$ that has the largest gradient
      \IF{$\SaniLoss < \MINLOSS$}
      \STATE $\MINLOSS \leftarrow \SaniLoss$ 
      \STATE $\OptimalSanitizedAdj \leftarrow \SanitizedAdj$ \COMMENTS{Save the optimal adjacency matrix.}
      \ENDIF
      \ELSE
      \STATE \textbf{break}  \COMMENTS{There is no edge to prune.}
      \ENDIF
     \STATE $n \leftarrow n + 1$
     \ENDWHILE \\
    \RETURN Sanitized Graph $\SanitizedGraph = (\OptimalSanitizedAdj, \NodeFeature)$
    \end{algorithmic}
\end{algorithm}
\end{figure}
\prop\ takes a poisoned graph $\PoisonedGraph = (\PoisonedAdj, \NodeFeature)$ as input.
First of all, $\SanitizedAdj$ is initialized as the poisoned one~$\PoisonedAdj$ (Line~2).
Similarly, the optimal adjacency matrix $\OptimalSanitizedAdj$is initialized as $\SanitizedAdj$ (Line~3).
Note that $\OptimalSanitizedAdj$ is the output of Algorithm~\ref{alg1}.
Furthermore, the minimum loss $\MINLOSS$ is initialized (Line~4).
The encoder $\Encoder$ is trained with $\SanitizedAdj$ and $\NodeFeature$ (Line~6).
After that, depending on an assumed GCL method, $\AugGraph{m} = \ViewK{m}$ are obtained via the $k$-th stochastic augmentation, which is repeated $K$ times (Line 8).
By inputting $\AugGraph{m} = \ViewK{m}$ into the trained $\Encoder_{\Parameters^{'}}$, $\GradientSum$ is calculated (Line~13).
The candidates of pruned edges are selected from $\SanitizedAdj$ 
depending on whether they meet the two conditions in Eq.~(\ref{eq:del_condition}) and Eq.~(\ref{eq:fsim_condition}) (Line 15).
Let $\Candidate$ the candidate edges to prune.
Then, out of edges in $\Candidate$, \prop\ prunes one edge that has the largest gradient (Line 17).
Note that \prop\ executes only deleting edges although it is desirable that sanitizing poisoned graphs be conducted by edge modification including both deletion and addition of edges.
This is because \prop\ simply tries to solve optimization problems in Eq.~(\ref{eq:opt_problem1}) or Eq.~(\ref{eq:opt_problem2}) in a greedy manner at this stage.
If there is no edge in $\Candidate$, our edge pruning is stopped at that point.
Once an edge is deleted, $\SaniLoss$ is less than $\MINLOSS$, $\SanitizedAdj$ is saved as $\OptimalSanitizedAdj$ (Line~20).
After that, \prop\ tries to prune another edge in the next iteration by repeating the above mentioned procedures (Lines 5-26).
Finally, after up to $N$ edges are pruned, the sanitized graph $\SanitizedGraph = (\OptimalSanitizedAdj, \NodeFeature)$ is returned (Line~27).

\section{Experiments} \label{sec:experiment}
In this section, we evaluate the effectiveness of \prop.
In particular, we mainly evaluate the performance of the node classification on both the graphs sanitized by the \prop\ and the poisoned ones to reveal the following questions:
\begin{enumerate}
  \item Can \prop\ eliminate the adverse effects from the poisoned graphs by pruning edges?~\label{question:1}
  \item Is the feature similarity effective in selecting adversarial edges to prune on various datasets? \label{question:2}
  \item How does \prop\ affect the quality of the embeddings on clean graphs? \label{question:3}
  \item How does the number of pruned edges affect the quality of the embeddings on poisoned and clean graphs? \label{question:4}
  \item Is \prop\ effective against an adaptive attack? \label{question:5}
\end{enumerate}
Evaluation metric is the accuracy that is defined as the percentage of correctly predicted nodes to all the testing nodes.
The higher the accuracy is, the better the quality of embeddings is.
All our experiments are conducted on the NVIDIA GeForce RTX 3090 GPU with a 24GB memory.

\subsection{Experimental Setups}
\noindent\textbf{Dataset.}
We utilize the six datasets including Cora, CiteSeer \cite{yang2016revisiting}, Amazon-Computers, Amazon-Photo, Coauthor-CS, and Coauthor-Physics \cite{shchur2018pitfalls} following \cite{zhang2022unsupervised,feng2022adversarial}.
The details about these datasets are shown in Appendix.\ref{appendix:dataset}

\noindent\textbf{GCL.}
As for GCL methods, GCA \cite{zhu2021graph} and ARIEL \cite{feng2022adversarial} are utilized for training the encoder that outputs embeddings.
The details about GCL methods are in Appendix~\ref{appendix:GCL_setup}.

\noindent\textbf{Poisoning attack.}
We utilize CLGA \cite{zhang2022unsupervised} as the attack method in our experiments.
The details about the setup of CLGA are in Appendix~\ref{appendix:CLGA_setup}.

\noindent\textbf{Downstream task.}
The downstream task in our experiments is node classification.
We follow the evaluation scheme conducted in the other studies \cite{zhu2021graph,zhang2022unsupervised,feng2022adversarial}, where a single graph of each dataset is firstly trained by the GCL method, then the resulting embeddings are utilized to train and test a classifier for the node classification.
The details about the setup of downstream task are in Appendix~\ref{appendix:downstream_setup}.

\noindent\textbf{Baseline.}
We compare \prop\ with baselines to clarify the effectiveness of \prop.
We compare \prop\ to two baselines for supervised GNN, namely GNNGuard \cite{zhang2020gnnguard} and GNNJaccard \cite{wu2019adversarial}.
Although these baselines are not for GCL methods, this comparison clarifies to what extent our method can eliminate adverse effects from poisoned graphs compared to them. 
Detailed hyperparameters are described in Appendix~\ref{appendix:GNN_baselines}.
Furthermore, we consider a method that prunes multiple edges once at all as the baseline in this experiment.
As with \prop, the baseline selects edges to prune on the basis of the gradients of the contrastive losses. 
The difference from \prop\ is that the baseline always prunes a designated number of edges once at all.
In this experiment, we set the number of pruned edges at 10\% of all the edges in a graph.
There are the four baselines depending on the combinations of baseline and GCL methods, namely \baselinenfs(G), \baselinenfs(A), \baseline(A), and \baseline(G).
Note that G and A are the abbreviation of GCA and ARIEL, respectively.
As with \prop, \baselinenfs(G), \baselinenfs(A) do not utilize the feature similarity of neighboring nodes.
On the other hand, \baseline(A) and \baseline(G) select the edge to prune with the feature similarity.

\noindent\textbf{\prop.}
In \prop, the encoder is retrained on updated graphs by a supposed GCL method during the pruning procedure for one epoch.
This is because it is possible that training poisoned graphs for a large number of epochs makes the encoder poisoned, which causes inappropriate selection of pruned edges.
The number of stochastic augmentations $K$ shown in Line 8 in Algorithm~\ref{alg1} is set at 10 for calculating $\GradientSum$ in all experiments.
Since \prop\ is designed for eliminating adverse effects from poisoned graphs, \prop\ can be with any GCL methods including GCA and ARIEL. 
We define the proposed methods as the GCL methods that take sanitized graphs instead of poisoned ones.
There are the four proposed methods, specifically \propnfs\ (G), \propnfs\ (A), \prop\ (G), and \prop\ (A).
All the four proposals are allowed to prune up to 10\% of edges in a graph so that the contrastive losses of supposed GCL methods are minimized. 
The details bout \prop\ are shown in Appendix~\ref{appendix:EdgePruner_setup}.

\subsection{Effectiveness of Edge Pruning on Poisoned Graphs} \label{sec:eval_poisoned}
\begin{table*}[t]
  \caption{Accuracy $\pm$ standard deviation (\%) of node classification under CLGA. The bold and underlined accuracies are the best and the second best ones, respectively.}
  \label{tab:poisoned_results}
  \centering
  \begin{tabular}{c|c|c|c|c|c|c}
    \toprule
    \textbf{Method} & \textbf{Cora} & \textbf{CiteSeer} & \textbf{Amazon-Computers} & \textbf{Amazon-Photo} & \textbf{Coauthor-CS} & \textbf{Coauthor-Physics} \\
    \midrule
    \prop\ (A) & 80.91 $\pm$ 1.29 & \textbf{71.83 $\pm$ 0.69} & \underline{83.30 $\pm$ 0.74} & 88.37 $\pm$ 0.32 & \underline{89.13 $\pm$ 0.69} & 92.34 $\pm$ 0.51 \\ 
    \prop\ (G) & \textbf{81.70 $\pm$ 0.99} & 69.82 $\pm$ 1.83 & 81.17 $\pm$ 0.60 & 83.63 $\pm$ 0.42 & \textbf{90.90 $\pm$ 0.67} & \textbf{93.04 $\pm$ 0.38} \\ 
    \propnfs\ (A) & \underline{81.44 $\pm$ 1.10} & 69.80 $\pm$ 0.95 & \textbf{84.92 $\pm$ 0.63} & \underline{88.58 $\pm$ 0.33} & 88.14 $\pm$ 0.61 & 91.76 $\pm$ 0.56 \\ 
    \propnfs\ (G) & 79.83 $\pm$ 0.96 & 63.26 $\pm$ 1.27 & 82.63 $\pm$ 0.69 & 83.88 $\pm$ 0.55 & 85.35 $\pm$ 0.81 & 90.51 $\pm$ 0.31 \\ 
    \midrule
    \baseline\ (A) & 81.08 $\pm$ 0.67 & \underline{71.18 $\pm$ 0.75} & 83.19 $\pm$ 0.41 & \textbf{88.60 $\pm$ 0.39} & 86.73 $\pm$ 0.59 & 90.20 $\pm$ 0.44 \\
    \baseline\ (G) & 80.89 $\pm$ 1.01 & 63.17 $\pm$ 1.48 & 80.53 $\pm$ 0.58 & 82.53 $\pm$ 0.67 & 82.32 $\pm$ 0.50 & 89.48 $\pm$ 0.41 \\
    \baselinenfs\ (A) & 80.23 $\pm$ 0.99 & 70.24 $\pm$ 0.93 & 82.49 $\pm$ 1.01 & 88.36 $\pm$ 0.38 & 86.24 $\pm$ 0.52 & 89.50 $\pm$ 0.50 \\
    \baselinenfs\ (G) & 79.16 $\pm$ 1.09 & 61.56 $\pm$ 1.35 & 80.64 $\pm$ 0.69 & 85.71 $\pm$ 0.54 & 81.40 $\pm$ 0.44 & 88.90 $\pm$ 0.38 \\
    \midrule
    GNNGuard & 79.53 $\pm$ 1.49 & 69.93 $\pm$ 1.42 & 81.18 $\pm$ 0.75 & 87.57 $\pm$ 0.66 & 88.16 $\pm$ 0.69 & \underline{92.81 $\pm$ 0.32} \\
    GNNJaccard & 79.95 $\pm$ 0.73 & 67.46 $\pm$ 1.85 & 81.13 $\pm$ 0.97 & 87.99 $\pm$ 0.65 & 83.67 $\pm$ 0.86 & 89.93 $\pm$ 0.57 \\
    \midrule
    ARIEL & 80.77 $\pm$ 1.18 & 69.96 $\pm$ 0.94 & 82.51 $\pm$ 0.70 & 87.26 $\pm$ 0.37 & 85.35 $\pm$ 0.79 & 87.70 $\pm$ 0.38 \\
    GCA & 80.37 $\pm$ 0.79 & 61.73 $\pm$ 1.47 & 79.42 $\pm$ 0.77& 81.10 $\pm$ 0.47 & 81.30 $\pm$ 0.68 & 88.66 $\pm$ 0.35\\
  \bottomrule
\end{tabular}
\end{table*}
We evaluate the effectiveness of the proposed variants under CLGA by comparing the node classification accuracy on sanitized graphs with that on poisoned ones.
\tablename~\ref{tab:poisoned_results} shows the node classification accuracy under CLGA.
As shown in \tablename~\ref{tab:poisoned_results}, the variants of \prop\ achieve both the best and the second best accuracies on Cora, Amazon-Computers, Coauthor-CS, and Coauthor-Physics.
In the following discussions, we compare \prop\ with the other methods while referring to the results in \tablename~\ref{tab:poisoned_results}.

\smallskip
\noindent \textbf{Comparison with baselines.}
Compared with the our \baseline\ methods, the variants of \prop\ attains the best accuracies on the five datasets except for Amazon-Photo.
\prop\ (A) and \propnfs\ (A) achieve the best accuracies, namely 71.83\% and 84.92\% on CiteSeer and Amazon-Computers, respectively.
\prop\ (G) also yields the best accuracies on Cora, Coauthor-CS, and Coauthor-Physics.
On the other hand, \baseline\ (A) attains the best accuracy, 88.60\% only on Amazon-Photo.
As for other cases, the baselines are outperformed by \prop.
In particular, the accuracy of \baseline\ (G) on Coauthor-CS is 82.32\%, which is 8.58\% lower than that of \prop\ (G). 

Additionally, we consider that \prop\ is more competent than the \baseline\ variants from the point of view of robustness against an adaptive attack.
Since \baseline\ variants always prune a designated number of edges, they are vulnerable to an adaptive attack. 
In other words, it is possible for the adaptive attack to perturb edges so that the number of perturbed edges is larger than the number of pruned edges if the number of edges pruned by the \baseline\ method is leaked to the attacker.
On the other hand, \prop\ is immune to the adaptive attack.
We will show the details of the results on the adaptive attack in Section~\ref{sec:adaptive_attack}.
As for existing baselines for GNN, GNNGuard only achieves the second best accuracy on Coauthor-Physics.
Overall, GNNGuard and GNNJaccard are inferior to \prop\ although they counter poisoned graphs by using labels.
For these reasons, \prop\ is superior to the baselines on poisoned graphs.

\smallskip
\noindent \textbf{Comparison with existing GCL methods.}
Compared with GCA, \prop\ (G) outperforms GCA on all the datasets.
In particular, \prop\ (G) achieves 90.90\% accuracy on Coauthor-CS, which is 9.60\% higher than that of GCA.
Additionally, \prop\ (G) achieves 93.04\% accuracy on Coauthor-Physics, which is 4.38\% higher than that of GCA.
On the other hand, \prop\ (A) and \propnfs\ (A) also improve the accuracies in most cases compared with ARIEL.
In particular, \prop\ (A) yields 71.83\% accuracy on CiteSeer, which is the best accuracy.
In addition to that, \prop\ (A) achieves the second best accuracies, namely 83.30\% and 89.13\% accuracies on Amazon-Computers and Coauthor-CS, respectively.
Similarly, \propnfs\ also yields the best accuracy on Amazon-Computers.
The overall results demonstrate that \prop\ is effective in improving the quality of the embeddings.
Additionally, we consider that \prop\ may also be effective in sanitizing poisoned graphs created by other attacks on GCL as long as poisoned graphs are created by mainly adding adversarial edges. 
Thus, we conclude that \prop\ can eliminate adverse effects from the poisoned graphs, which answers the research question~(\ref{question:1}).

\smallskip
\noindent \textbf{Impact of similarity of node features.}
We compare \prop\ with \propnfs\ to evaluate the impact of feature similarity of neighboring nodes.
As shown in \tablename~\ref{tab:poisoned_results}, \prop\ (G) and \prop\ (A) do not always outperform \propnfs\ (G) and \propnfs\ (A).
In particular, \propnfs\ (G) achieves higher accuracies on Amazon-Computers and Amazon-Photo datasets compared with \prop\ (G).
Similarly, \propnfs\ (A) also outperforms \prop\ (A) on the two datasets.
Furthermore, \propnfs\ (A) yields 84.92\% accuracy on Amazon-Computers and 88.58\% accuracy on Amazon-Photo, which are the best and the second best ones, respectively.
These results mean that utilizing the feature similarity is not always required for sanitizing poisoned graphs.
The reason for this could be that the distributions of the cosine similarity of nodes connected by adversarial edges on Amazon-Photo and Amazon-Computers are different from those on the other datasets.
As shown in \figurename~\ref{fig:append_cos_hist} in Appendix~\ref{appendix:cos_hist}, the cosine similarity scores on the Amazon datasets are widely distributed.
This is why it may be difficult to identify adversarial edges on the two datasets compared with adversarial edges on other datasets.
However, compared with ARIEL, \propnfs\ (A) cannot improve the accuracies on CiteSeer.
\propnfs\ (G) also degrades the accuracy on Cora compared with GCA.
On the other hand, \prop\ (G) and \prop\ (A) can improve the accuracies of node classification compared with those of GCA and ARIEL, respectively, on all the datasets.
According to these results, we conclude that feature similarity is basically effective in selecting pruned edges on various datasets, which is the answer to the research question~(\ref{question:2}).

\subsection{Effect of Edge Pruning on Clean Graphs} \label{sec:clean_graph}
\begin{table*}[t]
  \caption{Accuracy $\pm$ standard deviation (\%) of node classification on clean graphs. The bold and underlined accuracies are the best and the second best ones, respectively.}
  \label{tab:clean_results}
  \centering
  \begin{tabular}{c|c|c|c|c|c|c}
    \toprule
    \textbf{Method} & \textbf{Cora} & \textbf{CiteSeer} & \textbf{Amazon-Computers} & \textbf{Amazon-Photo} & \textbf{Coauthor-CS} & \textbf{Coauthor-Physics} \\
    \midrule
    \prop\ (A) & 83.09 $\pm$ 0.95 & 72.16 $\pm$ 0.87 & 85.92 $\pm$ 0.73 & 91.02 $\pm$ 0.28 & \textbf{90.13 $\pm$ 0.55} & 93.10 $\pm$ 0.39 \\ 
    \prop\ (G) & 81.96 $\pm$ 1.18 & 70.29 $\pm$ 1.62 & 83.92 $\pm$ 0.63 & 90.39 $\pm$ 0.44 & \underline{90.12 $\pm$ 0.60} & \textbf{93.52 $\pm$ 0.33}  \\ 
    \propnfs\ (A) & 83.40 $\pm$ 0.57& \underline{72.55 $\pm$ 0.72} & 85.77 $\pm$ 0.64 & 91.03 $\pm$ 0.47 & 89.87 $\pm$ 0.55 & 92.74 $\pm$ 0.59\\ 
    \propnfs\ (G)  & 81.58 $\pm$ 0.98 &	69.64 $\pm$ 1.74 & 82.80 $\pm$ 0.74 & 89.81 $\pm$ 0.52 & 89.39 $\pm$ 0.43 & \underline{93.37 $\pm$ 0.27} \\ 
    \midrule
    \baseline\ (A) & 82.65 $\pm$ 0.75 & 72.25 $\pm$ 0.91 & \textbf{87.95 $\pm$ 0.58} & 91.93 $\pm$ 0.41 & 89.80 $\pm$ 0.63 & 92.83 $\pm$ 0.41 \\
    \baseline\ (G) & 82.73 $\pm$ 1.21 & 70.46 $\pm$ 1.32 & 85.61 $\pm$ 0.72 & 88.79 $\pm$ 0.43 & 89.15 $\pm$ 0.71 & 93.11 $\pm$ 0.30 \\
    \baselinenfs\ (A) & 82.73 $\pm$ 1.02 & 71.99 $\pm$ 0.67 & 87.20 $\pm$ 0.78 & \textbf{92.34 $\pm$ 0.36} & 89.39 $\pm$ 0.75 & 92.64 $\pm$ 0.30 \\
    \baselinenfs\ (G) & 82.99 $\pm$ 1.03 & 70.28 $\pm$ 1.64 & 86.20 $\pm$ 0.65 & 88.96 $\pm$ 0.38 & 88.85 $\pm$ 0.58 & 92.92 $\pm$ 0.32 \\
    \midrule
    GNNGuard & 79.69 $\pm$ 1.56 & 69.85 $\pm$ 1.36 & 85.96 $\pm$ 0.56 & 91.88 $\pm$ 0.54 & 88.69 $\pm$ 0.71 & 92.81 $\pm$ 0.44 \\
    GNNJaccard & 82.18 $\pm$ 0.62 & 70.41 $\pm$ 1.01 & 86.15 $\pm$ 0.43 & \underline{92.22 $\pm$ 0.31} & 87.47 $\pm$ 0.98 & 92.68 $\pm$ 0.46 \\
    \midrule
    ARIEL & \underline{83.97 $\pm$ 0.83} & \textbf{72.69 $\pm$ 0.74} & \underline{87.38 $\pm$ 0.70} & 92.08 $\pm$ 0.41 & 89.50 $\pm$ 0.57 & 92.52 $\pm$ 0.35 \\ 
    GCA & \textbf{84.12 $\pm$ 1.24} & 70.40 $\pm$ 1.38 & 86.27 $\pm$ 0.51 & 90.78 $\pm$ 0.45 & 88.87 $\pm$ 0.49 & 93.07 $\pm$ 0.39 \\
  \bottomrule
\end{tabular}
\end{table*}
In this subsection, we evaluate the effect of our edge pruning on clean graphs.
It is difficult for defenders to judge whether an input graph is clean or poisoned every time.
Thus, it is desirable that edge pruning is applicable to input graphs regardless of whether they are poisoned.
\tablename~\ref{tab:clean_results} shows the accuracy of node classification on clean graphs.
In \tablename~\ref{tab:clean_results}, we report the results of node classification evaluated on clean subgraphs of 5,000 nodes on Amazon-Computers, Amazon-Photo, Coauthor-CS, and Coauthor-Physics.
As shown in \tablename~\ref{tab:clean_results}, ARIEL basically outperforms GCA on all datasets except for Coauthor-Physics.

\smallskip
\noindent \textbf{Comparison with baselines.}
As we can see from \tablename~\ref{tab:clean_results}, the best and the second best accuracies appear in different methods depending on the datasets.
\baseline\ (A) and \baselinenfs\ (A) achieve the best accuracies on Amazon-Computers and Amazon-Photo, respectively.
On the other hand, \prop\ (A) and \prop\ (G) yield the best accuracies on Coauthor-CS and Coauthor-Physics, respectively.
Considering the best accuracies, the proposed methods are comparable to the our \baseline\ variants.
However, \prop\ (G) \propnfs\ (A) and \propnfs\ (G) yield the second best accuracies on Coauthor-CS, CiteSeer, and Coauthor-Physics, respectively.
On the other hand, there is no the second best accuracy in \baseline\ variants.
As for existing baselines for GNN, GNNJaccard only achieves the second best accuracy on Amazon-Photo.
Although GNNGuard and GNNJaccard achieve relatively better accuracies on Amazon datasets compared with \prop\ variants.
However, they are inferior to \prop\ on the other datasets.
From these results, the variants of \prop\ slightly outperforms the baselines on clean graphs.

\smallskip
\noindent \textbf{Comparison with existing GCL methods.}
In the case of Coauthor-CS and Coauthor-Physics, it is observed that \prop\ (A) and \prop\ (G) can yield better accuracies than ARIEL and GCA, respectively.
These results mean that our pruning is also capable of eliminating adverse edges in clean graphs.
To be precise, \prop\ (A), and \prop\ (G) achieve the best accuracies, namely 90.13\%, and 93.52\% accuracies on Coauthor-CS and Coauthor-Physics, respectively.
In addition to that, \prop\ (G), \propnfs\ (A), and \propnfs\ (G) also yield the second best accuracies on Coauthor-CS, CiteSeer, and Coautor-Physics, respectively.

On the other hand, both \prop\ (G) and \propnfs\ (G) are inferior to GCA on Cora, CiteSeer, Amazon-Computers, and Amazon-Photo.
Similarly, both \prop\ (A) and \propnfs\ (A) are inferior to ARIEL on these four datasets.
In the worst case, compared with GCA, \propnfs\ (G) degrades the accuracies on clean graphs by 2.54\%, 0.76\%, 3.47\%, and 0.97\% on Cora, CiteSeer, Amazon-Computers, and Amazon-Photo, respectively.
In other words, the accuracy drops are within approximately 3.5\%.
According to these results, it seems that our pruning may degrade the quality of embeddings on clean graphs depending on datasets, which is a limitation of \prop\ at this stage.
However, most importantly, when the same GCL method is utilized, the accuracies of the proposed methods on clean graphs are always high compared with the accuracies of GCA and ARIEL on poisoned graphs as shown in \tablename~\ref{tab:poisoned_results}, which means that applying \prop\ to clean graphs yields better situations than being attacked.
For example, \prop\ (G) achieves 90.39\% accuracy on clean graphs of Amazon-Photos, which is 9.29\% higher than the accuracy of GCA on the poisoned graph of Amazon-Photo.
For the other five datasets, \prop\ (G) on clean graphs outperforms GCA on poisoned graphs.
These results also apply to \prop\ (A).
The reason for this could be that deleting edges is less detrimental to the graph structure unlike attacks.

These results mean that \prop\ can raise lower bounds of accuracies on poisoned graphs while maintaining acceptable accuracies on clean graphs, which answers the research question~(\ref{question:3}).
Hence, \prop\ is worth utilizing regardless of whether input graphs are clean or poisoned because it prevents GCL methods from getting trapped into the worst case.

\subsection{Effect of Pruning Rates}
\begin{figure}[t]
    \scalebox{0.75}{
      \begin{tabular}{cc}
        \subfigure[Cora]{
          \includegraphics[scale=0.35]{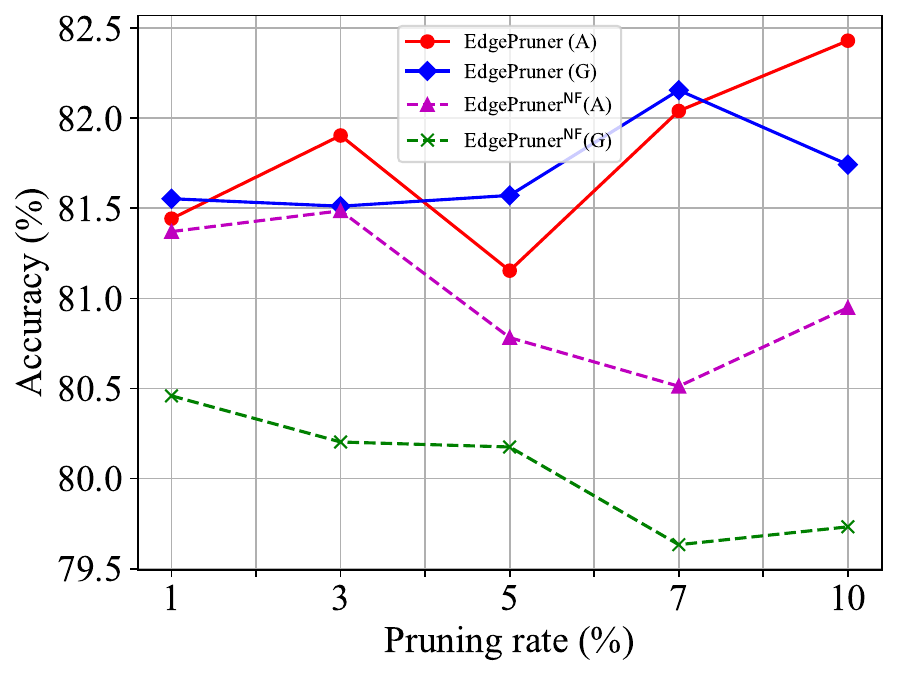}
          \label{fig:cora_acc_pr}
        } & \hspace{-0.7cm}
        \subfigure[CiteSeer]{
          \includegraphics[scale=0.35]{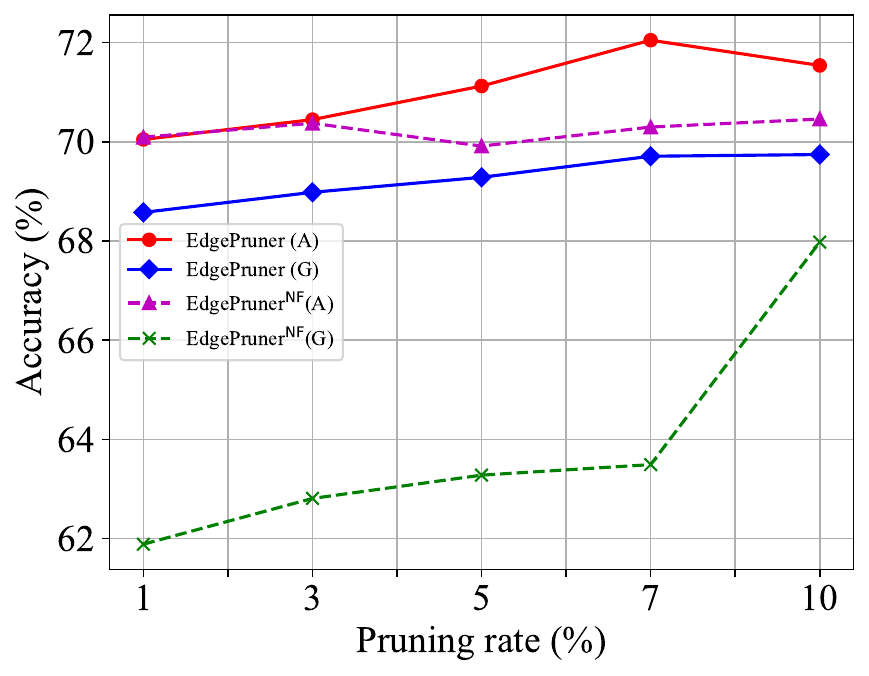}
          \label{fig:citeseer_acc_pr}
        } \\ 
        \subfigure[Amazon-Computers]{
          \includegraphics[scale=0.35]{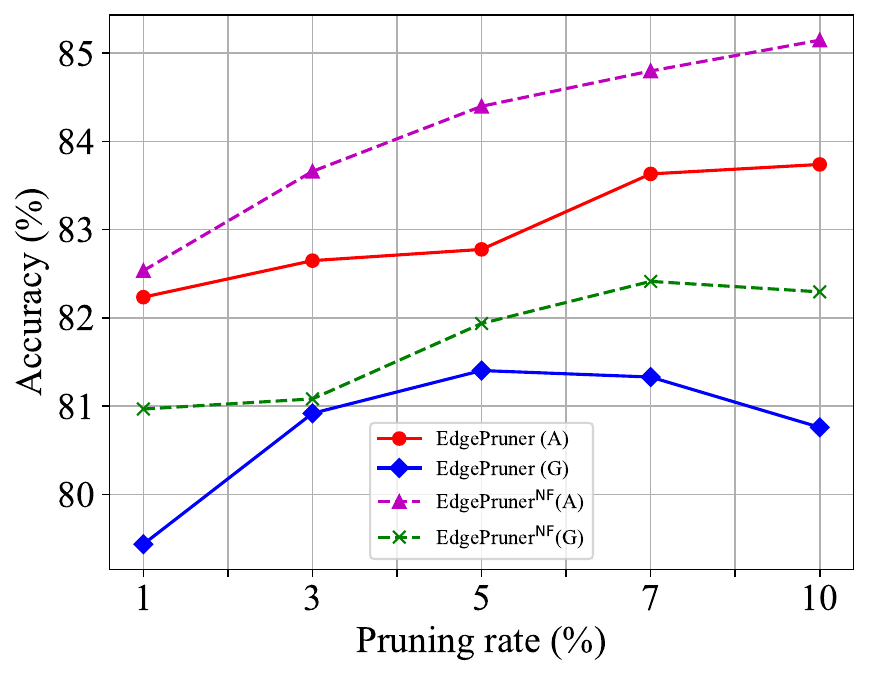}
          \label{fig:amazon_computers_acc_pr}
        } & \hspace{-0.7cm}
        \subfigure[Amazon-Photo]{
          \includegraphics[scale=0.35]{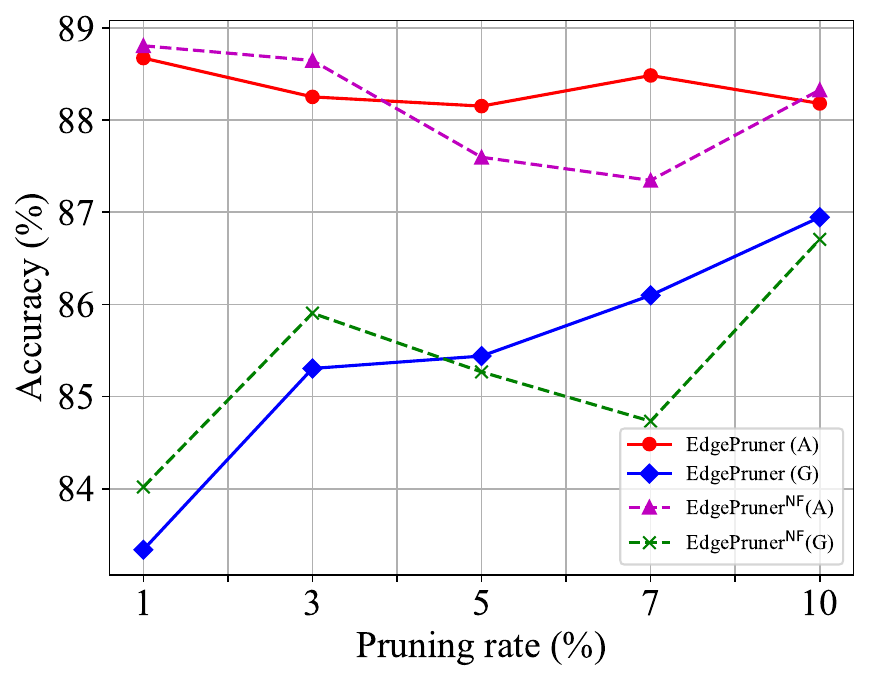}
          \label{fig:amazon_photo_acc_pr}
        } \\ 
        \subfigure[Coauthor-CS]{
          \includegraphics[scale=0.35]{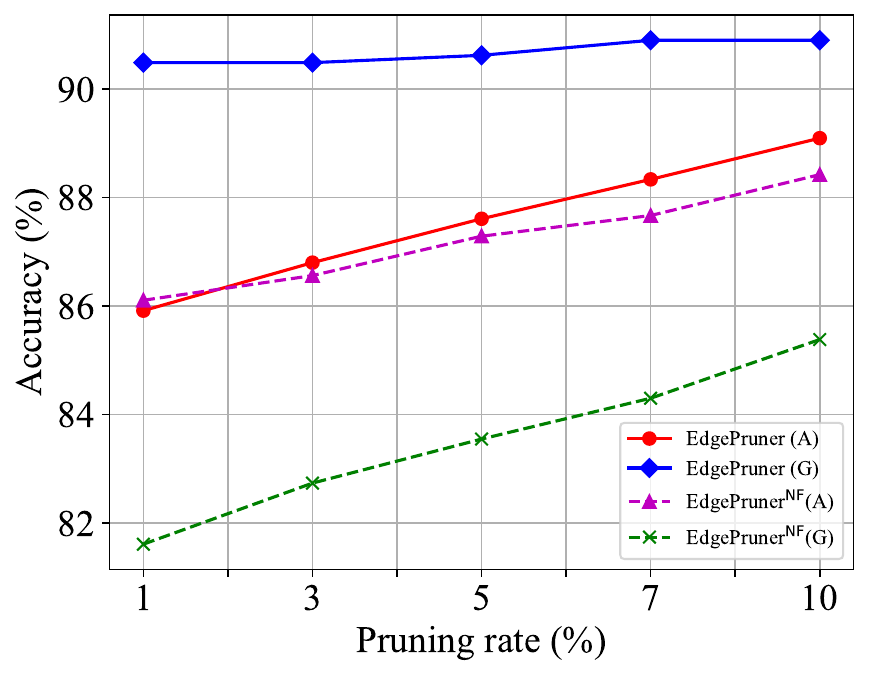}
          \label{fig:coauthor_cs_acc_pr}
        } & \hspace{-0.7cm}
        \subfigure[Coauthor-Physics]{
          \includegraphics[scale=0.35]{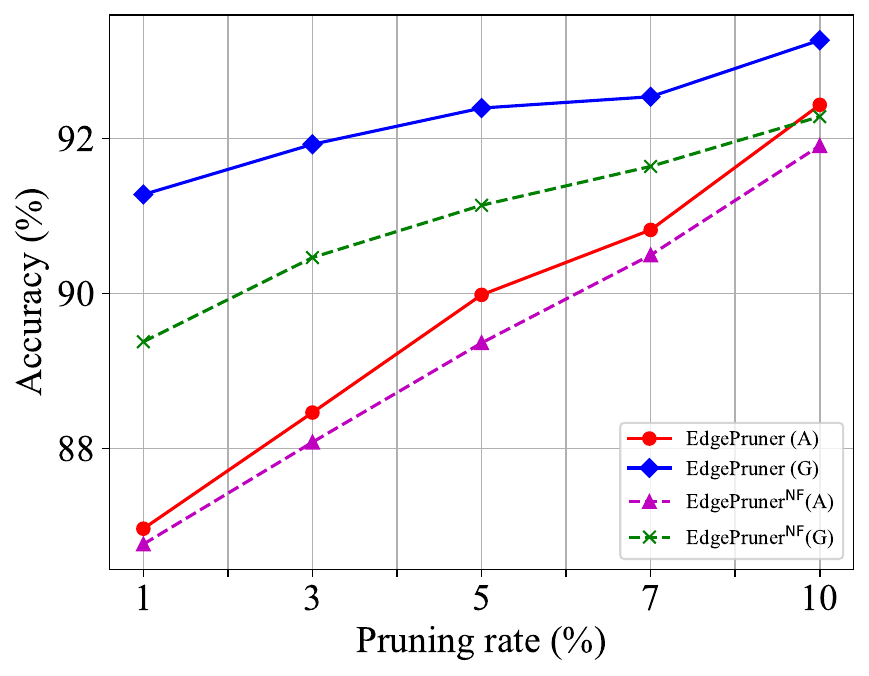}
          \label{fig:coauthor_physics_acc_pr}
        } \\
      \end{tabular}
    }
    \caption{Accuracies on poisoned graphs as a function of the pruning rate.}
    \label{fig:acc_pr}
\end{figure}

In this subsection, we evaluate the relationship between the number of pruned edges and the node classification accuracy.
\figurename~\ref{fig:acc_pr} shows the accuracy of node classification on poisoned graphs as a function of the pruning rate.
The pruning rate means the rate of deleted edges to all the edges in a graph.
In this experiment, we report the accuracies when pruning rates are 1\%, 3\%, 5\%, 7\%, and 10\%.
As shown in \figurename~\ref{fig:acc_pr}, as the pruning rate is increased, the accuracies are also increased in most cases.
\prop\ (A) tends to be always more competent than \propnfs\ (A) at most pruning rates except for Amazon-Photo and Amazon-Computers (comparing red lines with purple dotted ones). 
\prop\ (G) also has a similar tendency compared with \propnfs\ (G) (comparing blue lines with green dotted ones) as with \prop\ (A).
For example, as we can see from red and blue lines in \figurename~\ref{fig:coauthor_physics_acc_pr}, the accuracies of \prop\ (A) and \prop\ (G) on Coauthor-Physics are considerably improved as the pruning rates are increased.
In particular, the accuracy of \prop\ (A) on Coauthor-Physics is increased by approximately 5\% when the pruning rates change from 1\% to 10\%.
Furthermore, \prop\ (A) achieves the similar performance on Coauthor-CS as shown in \figurename~\ref{fig:coauthor_cs_acc_pr}.
On the other hand, as we can see from the green dotted line and the purple dotted one in \figurename~\ref{fig:cora_acc_pr}, the accuracies of \propnfs\ (A) and \propnfs\ (G) on Cora decrease gradually as the pruning rate increases whereas \prop\ (A) and \prop\ (G) finally improve accuracies when the pruning rate is 10\%.

\begin{table*}[t]
  \caption{Pruning rates (\%) when the final $\OptimalSanitizedAdj$ is obtained. (\prop\ is applied to poisoned graphs)}
  \label{tab:pruning_rate_optimal_graph_poison}
  \centering
      \begin{tabular}{c|c|c|c|c|c|c}
        \toprule
        \textbf{Method} & \textbf{Cora} & \textbf{CiteSeer} & \textbf{Amazon-Computers} & \textbf{Amazon-Photo} & \textbf{Coauthor-CS} & \textbf{Coauthor-Physics} \\
        \midrule
        \prop\ (A) &  1.89 & 7.26 & 9.87 & 3.23 & 9.69 & 9.85 \\ 
        \prop\ (G) &  10.00 & 7.03 & 9.78 & 0.53 & 8.68 & 9.53 \\ 
        \propnfs\ (A) &  7.77 & 0.73 & 9.26 & 3.23 & 8.31 & 9.85 \\
        \propnfs\ (G) &  7.35 & 0.05 & 8.05 & 0.02 & 0.10 & 3.35 \\ 
      \bottomrule
    \end{tabular}
\end{table*}

In the following, we discuss the meaning of our loss based graph selection utilized in \prop\ on the basis of the pruning rate.
\tablename~\ref{tab:pruning_rate_optimal_graph_poison} shows the pruning rates when the final $\OptimalSanitizedAdj$ is obtained in \prop\ on poisoned graphs.
As we can see from \tablename~\ref{tab:pruning_rate_optimal_graph_poison}, pruning rates are different depending on the variants and datasets.
In comparison with the pruning rates in \figurename~\ref{fig:acc_pr} and those in \tablename~\ref{tab:pruning_rate_optimal_graph_poison}, it is observed that the best variant on each dataset obtains the final $\OptimalSanitizedAdj$ around the pruning rates when better accuracies are achieved in \figurename~\ref{fig:acc_pr}. 
For example, \prop\ (A) obtains the final $\OptimalSanitizedAdj$ on CiteSeer when the pruning rate is 7.26\% as shown in \tablename~\ref{tab:pruning_rate_optimal_graph_poison}, which is almost consistent with the result that \prop\ (A) attains approximately 72\% accuracy at 7\% pruning rate as shown in \figurename~\ref{fig:citeseer_acc_pr}.
Similarly, in the case of Coauthor-Physics, \prop\ (G) gets the final $\OptimalSanitizedAdj$ when the pruning rate is 9.53\%, and the accuracy is the highest at 10\% pruning rate as shown in \figurename~\ref{fig:coauthor_physics_acc_pr}.
On the other hand, there are cases where the pruning rates are inconsistent.
For example, \prop\ (A) obtains the final $\OptimalSanitizedAdj$ on Cora when the pruning rate is 1.89\% although the highest accuracy is yielded when the pruning rate is 10\% as shown in \figurename~\ref{fig:cora_acc_pr}.
This means that adopting graphs when the pruning rate is 10\% is better.

\begin{figure}[t]
    \scalebox{0.75}{
      \begin{tabular}{cc}
        \subfigure[Cora]{
          \includegraphics[scale=0.35]{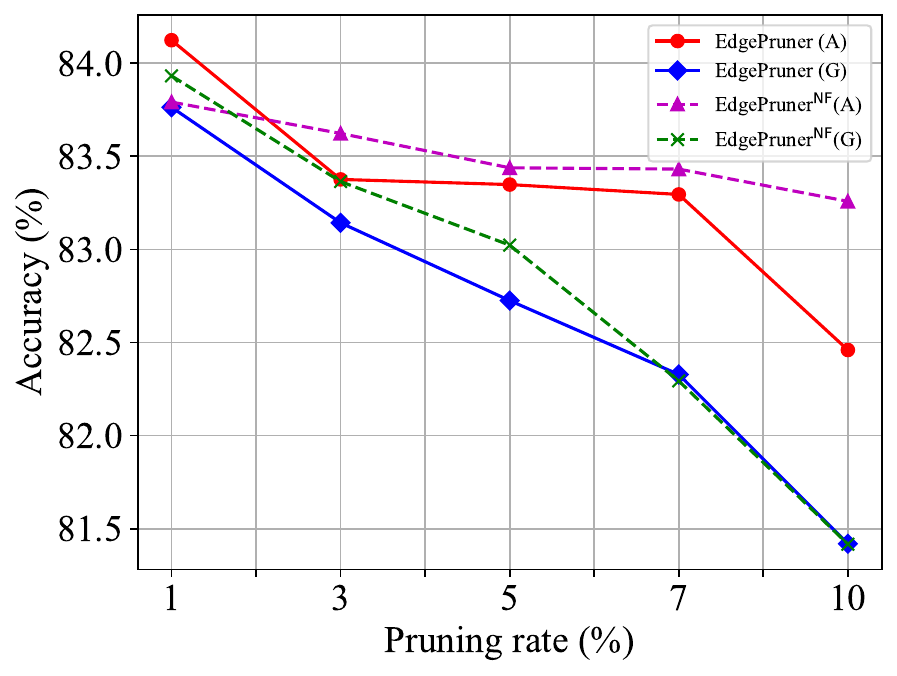}
          \label{fig:cora_acc_pr_clean}
        } & \hspace{-0.7cm}
        \subfigure[CiteSeer]{
          \includegraphics[scale=0.35]{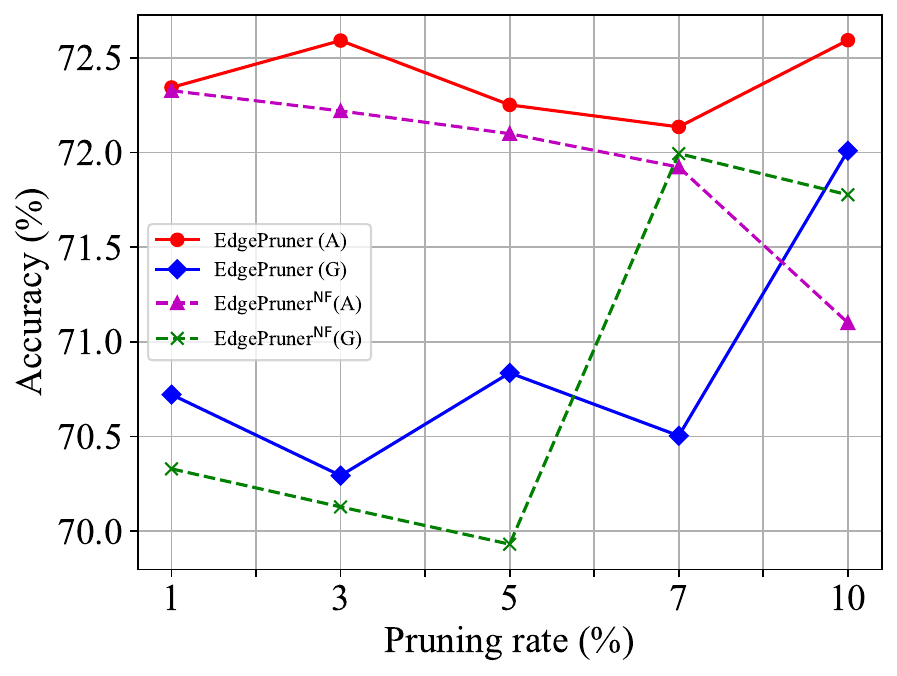}
          \label{fig:citeseer_acc_pr_clean}
        } \\ 
        \subfigure[Amazon-Computers]{
          \includegraphics[scale=0.35]{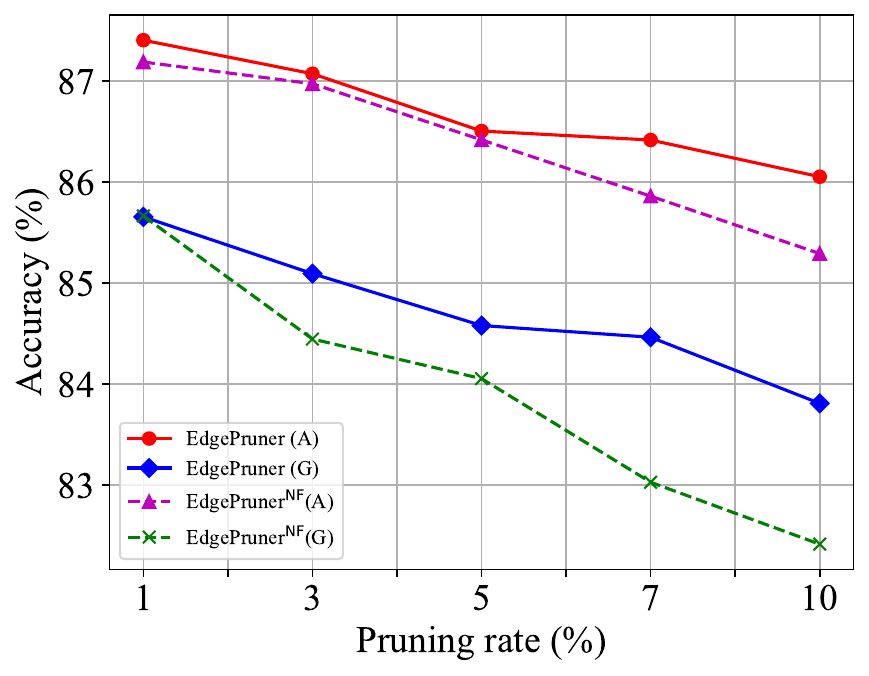}
          \label{fig:amazon_computers_acc_pr_clean}
        } & \hspace{-0.7cm}
        \subfigure[Amazon-Photo]{
          \includegraphics[scale=0.35]{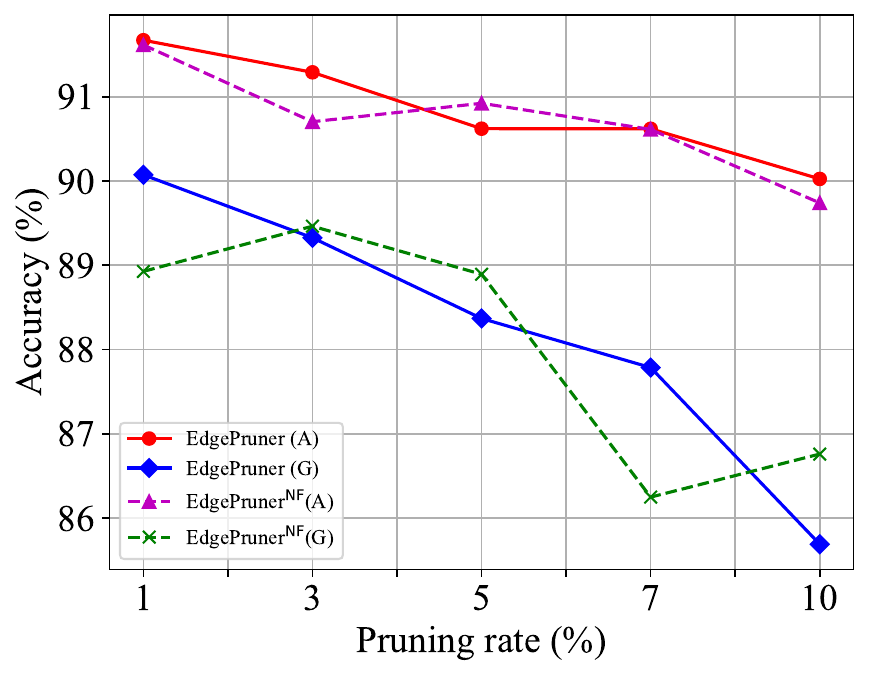}
          \label{fig:amazon_photo_acc_pr_clean}
        } \\ 
        \subfigure[Coauthor-CS]{
          \includegraphics[scale=0.35]{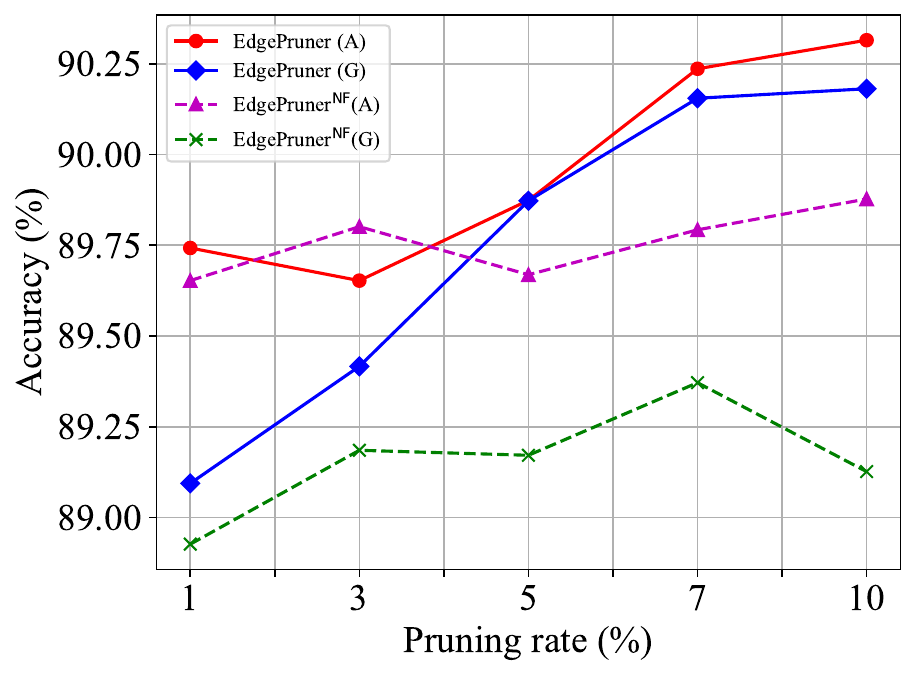}
          \label{fig:coauthor_cs_acc_pr_clean}
        } & \hspace{-0.7cm}
        \subfigure[Coauthor-Physics]{
          \includegraphics[scale=0.35]{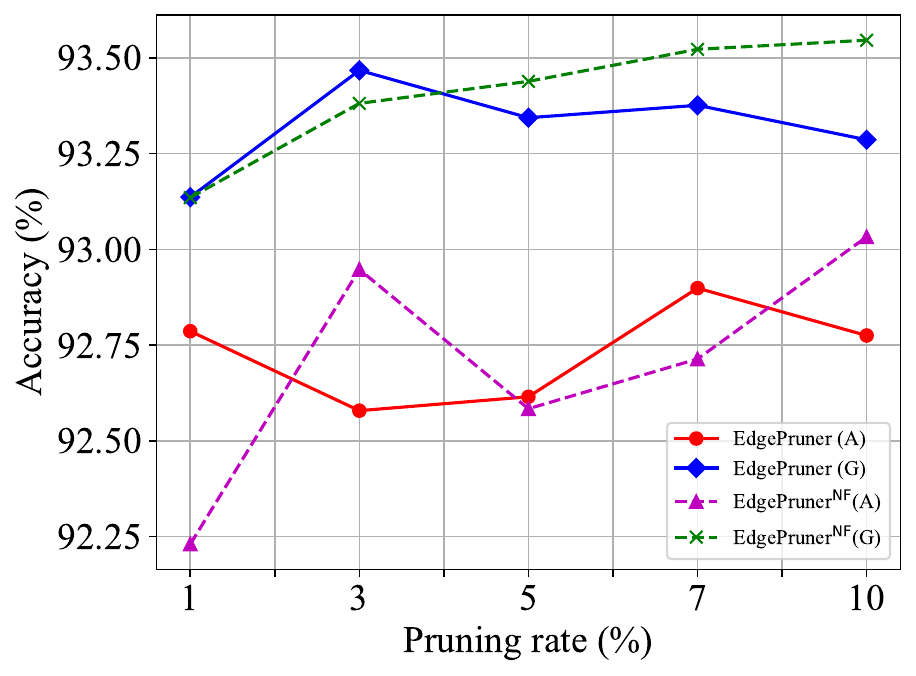}
          \label{fig:coauthor_physics_acc_pr_clean}
        } \\
      \end{tabular}
    }
    \caption{Accuracies on clean graphs as a function of the pruning rate.}
    \label{fig:acc_pr_clean}
\end{figure}

However, we consider that utilizing the minimum losses to select the final graphs is more effective. 
This is because this loss based graph selection is helpful in alleviating the bad influence on clean graphs.
We also evaluate the relationship between the number of pruned edges and the accuracies of node classification on clean graph.
\figurename~\ref{fig:acc_pr_clean} shows the accuracy of node classification on clean graphs as a function of the pruning rate.
As shown in \figurename~\ref{fig:acc_pr_clean}, the accuracies tend to be decreased as the pruning rate is increased.
In particular, the tendency is clearly observed on Cora, Amazon-Computers, and Amzoon-Photo as shown in \figurename~\ref{fig:cora_acc_pr_clean}, \figurename~\ref{fig:amazon_computers_acc_pr_clean}, and \figurename~\ref{fig:amazon_photo_acc_pr_clean}, respectively.
\tablename~\ref{tab:pruning_rate_optimal_graph_clean} shows the pruning rates when the final $\OptimalSanitizedAdj$ is obtained in \prop\ on clean graphs.
As shown in \tablename~\ref{tab:pruning_rate_optimal_graph_clean}, on Amazon-Photo, \prop\ (A) gets the final $\OptimalSanitizedAdj$ when the pruning rate is 3.54\%.
Similarly, \prop\ (A) obtains the final $\OptimalSanitizedAdj$ when the pruning rate is 6.56\% on Cora.
The accuracies around them are higher compared with those when the pruning rate is 10\% as shown in \figurename~\ref{fig:amazon_photo_acc_pr_clean}.
Meanwhile, as shown in \figurename~\ref{fig:coauthor_cs_acc_pr_clean}, the accuracies of \prop\ (A) and \prop\ (G) on the clean graph of Coauthor-CS are increased as the pruning rate is increased. 
These accuracies are the highest when the pruning rate is 10\%.
In such a dataset, \prop\ (A) and \prop\ (G) adopt the final $\OptimalSanitizedAdj$ at the pruning rates that are close to 10\% as shown in \tablename~\ref{tab:pruning_rate_optimal_graph_clean}.
These results mean that \prop\ can select better graphs as much as possible on the basis of the losses on clean graphs.
Thus, our loss based graph selection in \prop\ is more effective than adopting graphs after a designated number of edges are always pruned. 

From these results, we conclude that pruning more edges in poisoned graphs basically contributes to improvement to the quality of the embeddings although pruning too many edges in clean graphs may degrade the quality on clean graphs, which answers the research question~(\ref{question:4}).

\begin{table*}[t]
  \caption{Pruning rates (\%) when the final $\OptimalSanitizedAdj$ is obtained. (\prop\ is applied to clean graphs)}
  \label{tab:pruning_rate_optimal_graph_clean}
  \centering
      \begin{tabular}{c|c|c|c|c|c|c}
        \toprule
        \textbf{Method} & \textbf{Cora} & \textbf{CiteSeer} & \textbf{Amazon-Computers} & \textbf{Amazon-Photo} & \textbf{Coauthor-CS} & \textbf{Coauthor-Physics} \\
        \midrule
        \prop\ (A) &  6.56 & 3.71 & 8.99 & 3.54 & 8.53 & 4.21 \\ 
        \prop\ (G) &  8.09 & 5.91 & 8.71 & 0.07 & 8.16 & 3.68 \\ 
        \propnfs\ (A) &  6.56 & 0.09 & 7.39 & 3.54 & 8.14 & 9.89 \\
        \propnfs\ (G) &  9.45 & 2.77 & 7.94 & 0.07 & 8.16 & 3.68 \\ 
      \bottomrule
    \end{tabular}
\end{table*}

\subsection{Adaptive Attack} \label{sec:adaptive_attack}
\begin{table*}[t]
  \caption{Accuracy $\pm$ standard deviation (\%) of node classification under adaptive CLGA. The bold is the best accuracy.}
  \label{tab:adaptive_result}
  \centering
  \begin{tabular}{c|c|c|c|c|c|c}
    \toprule
    \textbf{Method} & \textbf{Cora} & \textbf{CiteSeer} & \textbf{Amazon-Computers} & \textbf{Amazon-Photo} & \textbf{Coauthor-CS} & \textbf{Coauthor-Physics} \\
    \midrule
    \prop\ (G) & \textbf{78.89 $\pm$} 0.78 & \textbf{63.92 $\pm$ 2.08} & \textbf{79.41 $\pm$ 0.62} & \textbf{83.45 $\pm$ 0.40} & \textbf{85.15 $\pm$ 0.60} & \textbf{88.54 $\pm$ 0.24} \\ 
    GCA & 77.22 $\pm$ 1.19 & 59.61 $\pm$ 1.85 & 75.68 $\pm$ 0.89 & 82.02 $\pm$ 0.68 & 78.58 $\pm$ 0.75 & 86.60 $\pm$ 0.47\\
  \bottomrule
\end{tabular}
\end{table*}
In this subsection, we evaluate the node classification accuracies under an adaptive attack against \prop.
We assume that an attacker knows that the pruning rate utilized in \prop\ is 10\%.
In the adaptive attack, poisoned graphs are created by modifying 15\% of edges in clean graphs in order to make it difficult to eliminate adverse effect by the poisoning attack.
We evaluate the node classification accuracies in the case where \prop(G) is in place.
As with other experiments, the proposed methods are allowed to prune up to 10\% of edges in the poisoned graphs.
\tablename~\ref{tab:adaptive_result} shows the node classification accuracy under adaptive CLGA.
As shown in \tablename~\ref{tab:adaptive_result}, \prop\ (G) improves node classification accuracy on all the datasets.
However, the accuracies are less improved compared with the situation of normal CLGA.
This is because pruning edges are not sufficient to eliminate the adverse effects of adaptive CLGA.
This result demonstrates that \prop\ is effective in sanitizing poisoned graphs created by the adaptive attack to some extent.

However, we believe that the adaptive attack against our defense is infeasible in the case where \prop\ is allowed to prune all the edges in a graph.
In other words, this case means that the pruning rate is set at 100\%.
This case is possible because \prop\ does not necessarily prune all the edges even in such a case.
\prop\ can flexibly select a sanitized graph on the basis of the minimum loss.
Furthermore, defenders who utilize \prop\ can also adopt another graph on the basis of the validation accuracy if they have a small amount of validation data consisting of labeled graphs.
Thus, in this case, an attacker cannot set a poisoning rate so that it is larger than the pruning rate utilized in \prop.
For these reasons, we conclude that \prop\ is effective against the adaptive attack from the perspectives of both our empirical results and the logic, which answers the research question~(\ref{question:5}).

\subsection{Discussion} \label{sec:discussion}

\smallskip
\noindent \textbf{Analysis of Pruned Edges in Poisoned Graph} 
Our method can improve the the node classification accuracies on poisoned graphs.
However, the accuracies are not completely improved to the level comparable to the ones on clean graphs.
We analyze the reason, and the results are in Appendix~\ref{appendix:pruned_edges_analysis}.

\smallskip
\noindent \textbf{Performance of Link Prediction}
We evaluate the performance of link prediction of proposed variants and the existing GCL methods on poisoned graphs. 
The detailed results are in Appendix~\ref{appendix:link_prediction}.

\smallskip
\noindent \textbf{The Effect of Edge Addition for Sanitizing Poisoned Graph}
In our experiments, it is observed that poisoned graphs against GCL are mainly created by adding poisoned edges in a graph.
This is why our experimental results show that our concept of pruning edges is effective.
However, edge deletion is also conducted when poisoned graphs on Amazon datasets are created. 
Thus, we conduct some experiments in order to confirm the effects of edge addition in the defense.
We evaluate the effect of edge addition for sanitizing poisoned graphs in the case where our method conducts both addition and deletion.
The detailed results are in Appendix~\ref{appedix:amazon_study}.

\section{Conclusions, Limitations, and Future work} \label{sec:conclusion}
In this paper, we have proposed \prop, \ti.
\prop\ prunes edges that contribute to minimizing the contrastive loss.
\prop\ can achieve more improvement to the quality of the embeddings under the poisoning attack.
Our experimental results demonstrate that pruning adversarial edges in poisoned graphs is feasible on the six datasets.
We demonstrate that the feature similarity of neighboring nodes is basically effective in determining pruned edges on various datasets.
Furthermore, \prop\ can also improve the quality of embedding even on the adaptive attack.

However, we have a main limitation, that is the fact that \prop\ cannot completely deal with the addition of edges, which may contribute to further improvement.
At this stage, \prop\ cannot sufficiently restore edges deleted by attacks.
Simply executing both the addition and the deletion of edges depending on situations is not so effective in our additional experiment.
However, we found that controlling edge addition is relatively promising way for defense.
We leave devising solutions for that to our future work.

Given the circumstances that there is no study that focuses on edge pruning in the GCL domain, we have shown its feasibility.
As a first step of the sanitization strategy against poisoning attacks on GCL, our work and experimental results can motivate researchers to design more effective defense methods for GCL.
To devise more practical countermeasures in the GCL domain, we plan to study how to more effectively restore clean graphs in the future.

\bibliographystyle{IEEEtran}
\bibliography{reference}

\appendix

\begin{table*}[t]
  \caption{The number of edges that are added and deleted when poisoned graphs are created on the basis of the loss of ARIEL.}
  \label{tab:adversarial_edges_ariel}
  \centering
  \begin{tabular}{c|c|c|c|c|c|c|c|c|c|c|c|c|c|c|c|c|c|c}
    \toprule
     & \multicolumn{3}{c|}{Cora} & \multicolumn{3}{c|}{CiteSeer} & \multicolumn{3}{c|}{Amazon-Computers} & \multicolumn{3}{c|}{Amazon-Photo}& \multicolumn{3}{c|}{Coauthor-CS} & \multicolumn{3}{c}{Coauthor-Physics} \\
    \cmidrule{2-19}
    & 1\% & 5\% & 10\% & 1\% & 5\% & 10\% & 1\% & 5\% & 10\% & 1\% & 5\% & 10\% & 1\% & 5\% & 10\% & 1\% & 5\% & 10\% \\
    \midrule
    Added edges & 52 & 263 & 528 & 29 & 106 & 216 & 338 & 1530 & 2869 & 496 & 2471 & 4898 & 61 & 309 & 619 & 52 & 258 & 519 \\
    Deleted edges & 0 & 0 & 0 & 16 & 121 & 239 & 18 & 253 & 699 & 0 & 9 & 62 & 0 & 0 & 0 & 0 & 4 & 6 \\
  \bottomrule
\end{tabular}
\end{table*}

\subsection{Overview of Graph Representation Learning} 
\subsubsection{Overview of GCN} \label{appendix:gcn}
In the GCN, to alleviate problems such as numerical instabilities and exploding or vanishing gradients during training, $\AdjMatrix$ is transformed into the symmetrically normalized adjacency matrix
\begin{equation}
  \hat{\AdjMatrix} = \tilde{\Diagonal}^{-\frac{1}{2}}\tilde{\AdjMatrix}\tilde{\Diagonal}^{-\frac{1}{2}},
\end{equation}
where $\hat{\AdjMatrix} = \AdjMatrix + \Identity$ is the adjacency matrix where self-connections are added to by the identity matrix $\Identity$.
$\tilde{\Diagonal}$ is the diagonal degree matrix of $\tilde{\AdjMatrix}$, which means $\Diagonal[i,j] = \sum_{j}{\hat{\AdjMatrix}[i,j]}$.
The encoder of the two-layer GCN is represented as 
\begin{equation}
  \Encoder(\AdjMatrix, \NodeFeature) = \sigma(\hat{\AdjMatrix}\sigma(\hat{\AdjMatrix}\NodeFeature \Weight^{(1)}) \Weight^{(2)}),
\end{equation}
where $\Weight^{(1)}$ and $\Weight^{(2)}$ are the trainable weights of the first and second layers, respectively.
Additionally, $\sigma$ is an activation function.

\subsubsection{Overview of GCA} \label{appendix:gca}
We introduce how to learn node embeddings following GCA model \cite{zhu2021graph}.
In a typical GCL, there are three steps to learn $\Encoder(\AdjMatrix, \NodeFeature)$, namely (1)~obtaining two views, (2)~obtaining node embeddings, and (3)~optimizing the parameters of $\Encoder(\AdjMatrix, \NodeFeature)$ on the basis of contrastive loss.
First of all, two augmented views $\View{1}$ and $\View{2}$ are generated from $\Graph$ through two stochastic augmentations $\Augmentation{1}$ and $\Augmentation{2}$, respectively.
Typical augmentation strategies include feature masking, edge dropping, and subgraph extraction, to name a few.
After obtaining the two views, they are fed into a shared encoder $\Encoder(\AdjMatrix, \NodeFeature)$ to obtain node embeddings.
Let $\EmbeddingMatrix{\Parameters}^{m}=\Encoder_{\Parameters}(\AdjMatrix_{m}, \NodeFeature_{m})$ denote the node embedding matrix transformed from $\Graph_{m}$ by $\Encoder$ with parameters $\Parameters$.
A contrastive loss is calculated for each node by using node embeddings in $\EmbeddingMatrix{\Parameters}^{m}$.
The contrastive loss is designed to gather similar nodes and to push dissimilar nodes away from each other.
In particular, the contrastive loss for the $i$-th node in the first view is calculated as 
\begin{equation}
  l(\EmbeddingOne, \EmbeddingTwo) = 
  -~\rm{log}\frac{\textit{e}^{\cossim{1}{i}{2}{i}/\tau}}
  {\textit{e}^{\cossim{1}{i}{2}{i}/\tau} + \sum_{\mathit{j \neq i}}{\textit{e}^{\cossim{1}{i}{1}{j}/\tau} + \textit{e}^{\cossim{1}{i}{2}{j}/\tau}}},
\end{equation}
where $\EmbeddingOne=\EmbeddingMatrix{\Parameters}^1[i,:]$ and $\EmbeddingTwo=\EmbeddingMatrix{\Parameters}^2[i,:]$ denote the embeddings of the $i$-th node in $\View{1}$ and $\View{2}$, respectively.
$\beta(\cdot)$ is a similarity function such as cosine similarity.
$\tau$ is a hyperparameter called temperature parameter.
Note that $l(\EmbeddingTwo, \EmbeddingOne)$ must also be calculated because the above loss is nonsymmetric for $\EmbeddingOne$ and $\EmbeddingTwo$.
Therefore, the final loss $\Loss$ is calculated as 
\begin{equation}
  \label{eq:loss}
  \Loss(\EmbeddingMatrix{\Parameters}^1, \EmbeddingMatrix{\Parameters}^2) = \frac{1}{2N} \sum_{i=1}^{N}{[l(\EmbeddingOne, \EmbeddingTwo) + l(\EmbeddingTwo, \EmbeddingOne)]}.
\end{equation}
Finally, $\Parameters$ are updated so that $\Loss(\cdot, \cdot)$ is minimized, which makes $\Encoder_\Parameters(\AdjMatrix, \NodeFeature)$ output similar node embeddings for similar nodes.

\subsection{Modified Edges in Poisoned Graphs for ARIEL} \label{appendix:num_edges_ariel}
\tablename~\ref{tab:adversarial_edges_ariel} shows the number of edges that modified when poisoned graphs are created on the basis of the loss of ARIEL.
The results are similar to the ones in the case where poisoned graphs are created on the basis of the loss of GCA.
The different tendency is observed on CiteSeer.
As we can see from \tablename~\ref{tab:adversarial_edges_ariel}, the number of deleted edges are increased and larger than the number of added edges on CiteSeer although there is no deleted edge on CiteSeer in \tablename~\ref{tab:adversarial_edges}.
We consider deleting edges is less effective than adding edges according to our experience.
This is why the attack performance for ARIEL on CiteSeer is not high in our experiment compared with the attack for GCA.

\subsection{Notation and Terminology for Our Algorithm}
\tablename~\ref{tab:notation} shows notations and their descriptions in Algorithm~\ref{alg1} for helping readers understand our method.

\begin{table*}[t]
  \caption{Notations and Terminologies used in Algorithm~\ref{alg1}.}
  \label{tab:notation}
  \centering
  \begin{tabular}{|c|l|}
    \hline
   \textbf{Notation} & \textbf{Description} \\ \hline
   $\PoisonedAdj$ & A poisoned adjacency matrix \\\hline
   $\NodeFeature$ & A node feature matrix \\\hline
   $\PoisonedGraph=(\PoisonedAdj, \NodeFeature)$ & A poisoned graph inputted into a GCL method \\\hline
   $\SanitizedAdj$ & A sanitized adjacency updated by pruning procedures \\\hline
   $\Encoder$ & An encoder trained by a GCL method \\ \hline
   $\SanitizedAdj^{k}_{m}$ & A $m$-th augmented sanitized adjacency matrix at the $k$-th augmentation out of $M$ views created by a GCL method \\ \hline
   $\NodeFeature^{k}_{m}$ & A $m$-th augmented node feature matrix at the $k$-th augmentation out of $M$ views created by a GCL method \\ \hline
   $\AugGraph{m} = \ViewK{m}$ & A $m$-th graph view obtained through $k$-th augmentation out of $M$ views created by a GCL method\\ \hline
   $\Parameters^{'}$ & Parameters of $\Encoder$ trained in a GCL method \\ \hline
   $\Gradient{m} = \frac{\partial \LossK}{\partial \AugAdj{m}}$ & A gradient matrix with regard to $\SanitizedAdj^{k}_{m}$ at the $k$-th augmentation \\ \hline
   $\nabla^{k} = \sum_{m}^{M}{\Gradient{m}}$  & A total gradient matrix obtained by adding up $\Gradient{m}$ at the $k$-th augmentation \\ \hline
   $\GradientSum$ & A total gradient matrix obtained by adding up $\nabla^{k}$ over $K$ times augmentations \\ \hline
   $\Candidate$ & Candidate edges to prune in each iteration \\ \hline
   $\SaniLoss$ & A contrastive loss of $\Encoder$ that trains $\SanitizedAdj$ \\ \hline
   $\MINLOSS$ & The minimum contrastive loss during our pruning procedures \\ \hline
   $\OptimalSanitizedAdj$ & A optimal adjacency matrix that produces the minimum contrastive loss \\ \hline
   $\SanitizedGraph = (\OptimalSanitizedAdj, \NodeFeature)$ & A sanitized graph consisting of $\OptimalSanitizedAdj$ and $\NodeFeature$ after pruning up to $N$ edges in a graph \\ \hline
   
    \hline
\end{tabular}
\end{table*}

\subsection{Dataset} \label{appendix:dataset}
\tablename~\ref{tab:dataset} shows the dataset summary.
These datasets are from PyTorch Geometric\footnote{All the datasets are from PyTorch Geometric 1.13.1 (https://pytorch-geometric.readthedocs.io/en/latest/modules/datasets.html)}.
Cora and CiteSeer \cite{yang2016revisiting} are citation networks where nodes and edges correspond to documents and their citations, respectively. 

Amazon-Computers and Amazon-Photo \cite{shchur2018pitfalls} are extracted from the copurchase graph in Amazon. 
In these graphs, nodes represent goods, and edges mean that two goods are frequently bought together.

Coauthor-CS and Coauthor-Physics \cite{shchur2018pitfalls} are the coauthorship graphs.
Nodes mean authors, and they are connected by an edge if there are the coauthorship between them on a paper.
\begin{table}[H]
  \caption{Summary of dataset used in our experiments. The number of nodes, edge, features, and classes are listed.}
  \label{tab:dataset}
  \centering
  \begin{tabular}{l|c|c|c|c}
    \toprule
    \textbf{Dataset name} & \textbf{Nodes} & \textbf{Edges} & \textbf{Features} & \textbf{Classes} \\
    \midrule
    Cora & 2,708 & 5,429 & 1,433 & 7  \\
    CiteSeer & 3,327 & 4,732 & 3,703 & 6\\
    Amazon-Computers & 13,752 & 245,861 & 767 & 10 \\
    Amazon-Photo & 7,650 & 119,081 & 745 & 8 \\
    Coauthor-CS & 18,333 & 81,894 & 6,805 & 15 \\
    Coauthor-Physics & 34,493 & 247,962 & 8,415 & 5 \\
    \bottomrule
  \end{tabular}
\end{table}

\subsection{Setup of GCL} \label{appendix:GCL_setup}
In both GCA and ARIEL, two-layer GCN is employed as the encoder as with the settings in GCA and ARIEL.
The implementation of GCA is based on the public implementation shared by the authors of CLGA because codes of GCA is also provided for evaluating the quality of the embeddings.
We also utilize the public implementations of ARIEL.
We adopt the hyperparameters used in the implementation of ARIEL, including the number of training epochs, the temperature parameter, edge dropping rates, and feature dropping rates in both GCA and ARIEL.
This is because utilizing the hyperparameters of ARIEL yields better node classification accuracy on clean graphs in our preliminary experiments.

\subsection{Setup of CLGA} \label{appendix:CLGA_setup}
In terms of four large datasets, namely Amazon-Computers, Amazon-Photo, Coauthor-CS, and Coauthor-Physics, we could not create poisoned graphs from their entire graphs by CLGA due to the memory limitations.
This is why we randomly sample a subgraph of 5,000 nodes for each of the above large graphs in accordance with the experimental setup in ARIEL.
The poisoned graphs are created on the basis of the subgraph.
We utilize the public implementation shared by the authors of CLGA in order to create poisoned graphs.
Poisoned graphs are created by modifying edges so that the contrastive loss of GCL methods is maximized.
In our experiments, it is assumed that GCA and ARIEL are attacked.
As for hyperparameters of GCA, except for the number of training epochs, we utilize the same hyperparameters as the ones used in the implementation of ARIEL when poisoned graphs are created.
We set the number of epochs for retraining at each iteration at one in our experiments because we can considerably reduce the computational cost while obtaining comparable attack performance to the results reported in \cite{zhang2022unsupervised} in our preliminary experiments.
To create the poisoned graphs, 10\% percent of edges in clean graphs are modified.

\subsection{Setup of Downstream Task} \label{appendix:downstream_setup}
We evaluate the quality of node embeddings produced by GCL methods.
Embeddings produced by GCA and ARIEL are utilized as features of the node classification.
The node classification is conducted by a logistic regression classifier that is implemented in scikit-learn \cite{sklean}.
When the node classification is evaluated, we split the nodes into 10\%, 10\%, and 80\% for training, validating, and testing, respectively.
As with the evaluation in \cite{feng2022adversarial}, the node classification is evaluated on 20 random dataset splits, which means that node classification is conducted 20 times on different sets. 
Thus, we report the mean testing accuracy of the node classification.

\subsection{Setup of \prop} \label{appendix:EdgePruner_setup}
\prop\ (G) and \prop\ (A) conduct edge pruning with the feature similarity of neighboring nodes.
We tentatively utilize values of $\Threshold$ that yield the best accuracy on the validation datasets in our preliminary experiments.
As a result, we set $\Threshold$ on Cora, CiteSeer, Amazon-Computers, Amazon-Photo, Coauthor-CS, and Coauthor-Physics at 0.10, 0.10, 0.30, 0.40, 0.25, and 0.25, respectively.
On the other hand, \propnfs\ (G) and \propnfs\ (A) do not utilize the feature similarity.
In the following experiments, we evaluate the effectiveness of the proposed methods by comparing them with the baselines and existing GCL methods, namely ARIEL and GCA.

\subsection{Distribution of Feature Similarity} \label{appendix:cos_hist}
\begin{figure*}[t]
    \centering
    \begin{tabular}{ccc}
      \subfigure[Cora]{
        \includegraphics[scale=0.33]{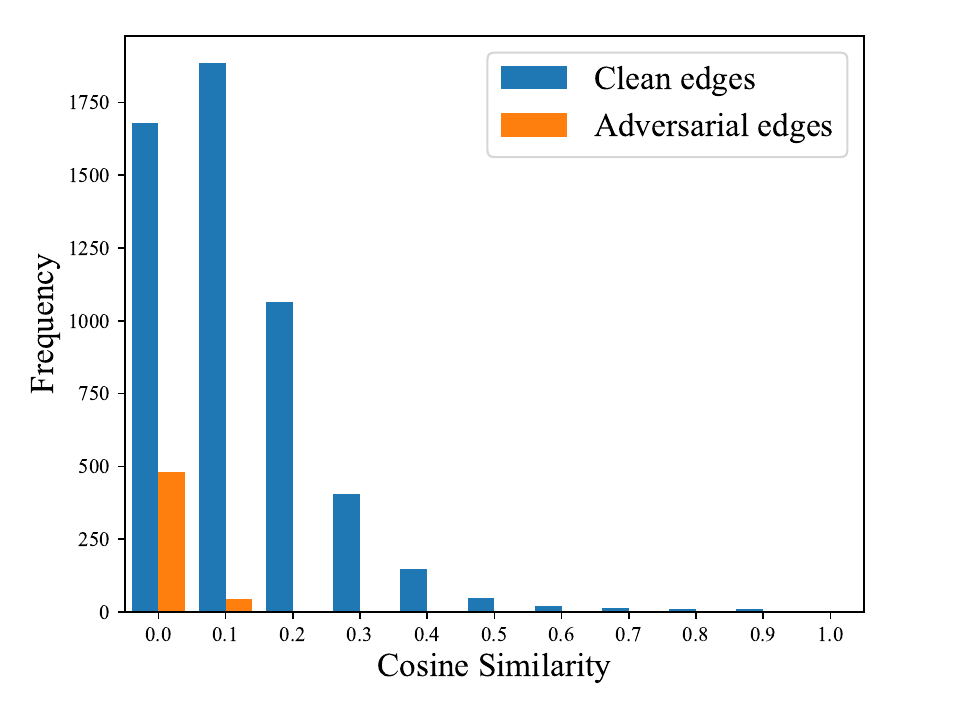}
        \label{fig:cora}
      } &
      \subfigure[CiteSeer]{
        \includegraphics[scale=0.33]{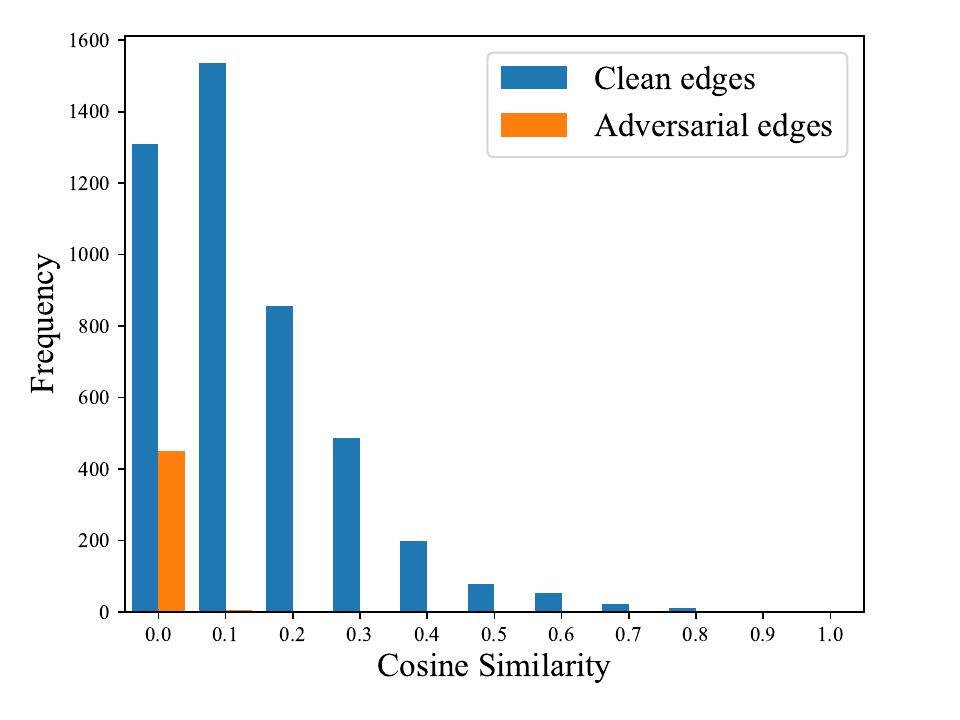}
        \label{fig:citeseer}
      } &
      \subfigure[Amazon-Computers]{
        \includegraphics[scale=0.33]{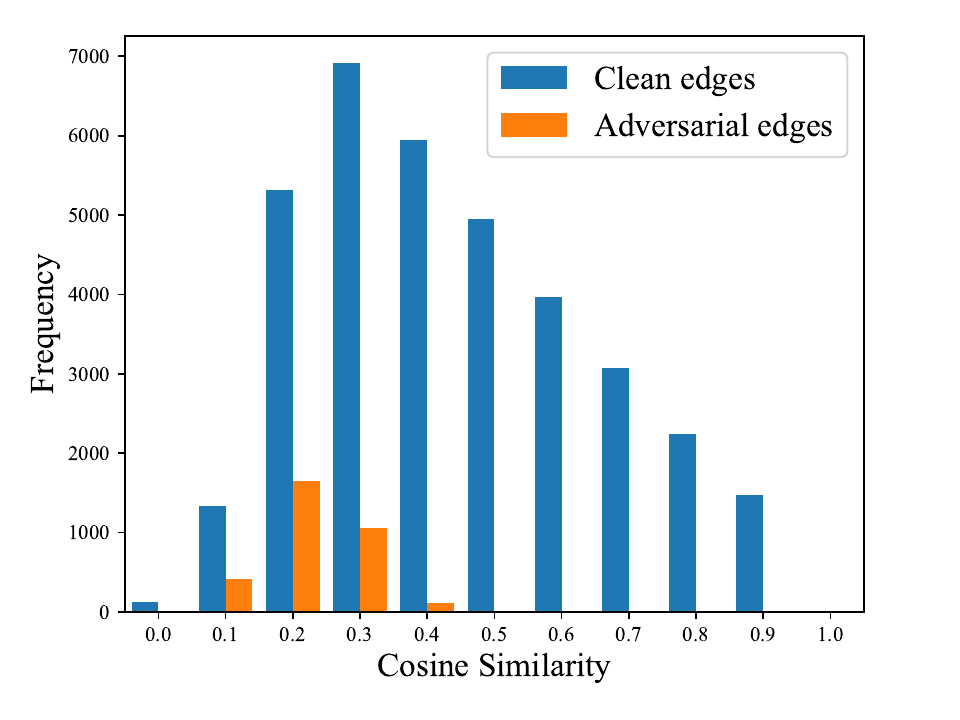}
        \label{fig:amazon_computers}
      } \\
      \subfigure[Amazon-Photo]{
        \includegraphics[scale=0.33]{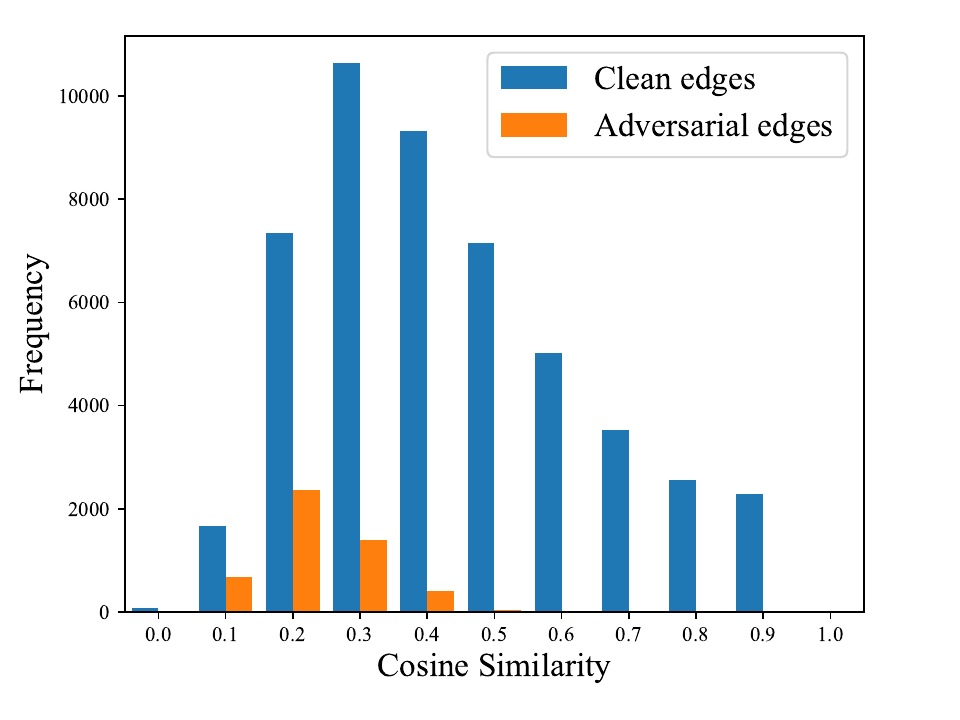}
        \label{fig:amazon_photo}
      }  &
      \subfigure[Coauthor-CS]{
        \includegraphics[scale=0.33]{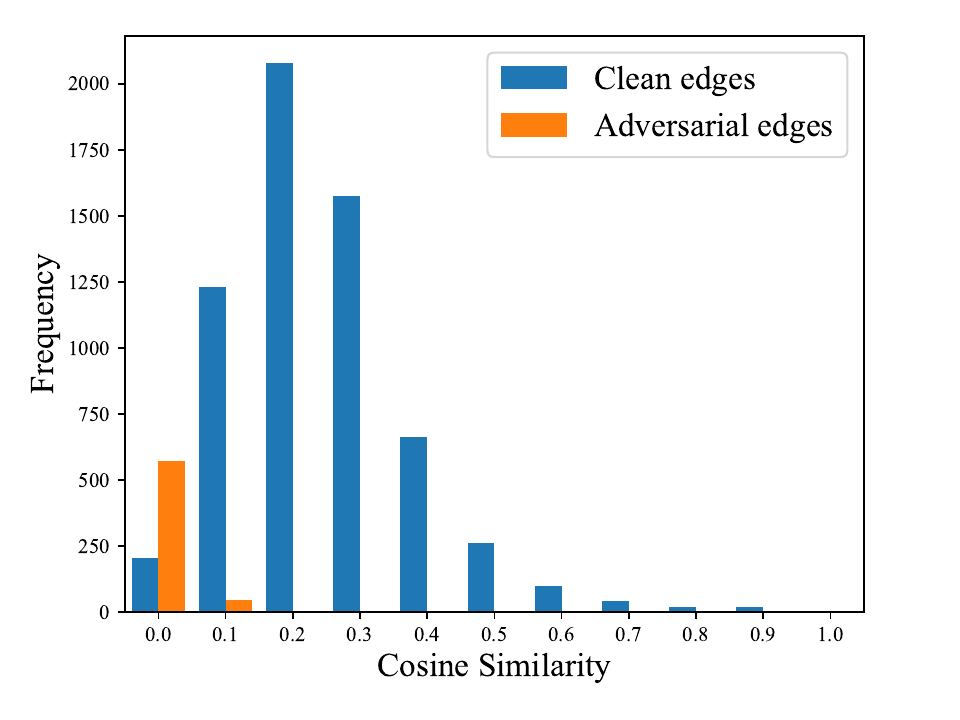}
        \label{fig:coauthor_cs}
      } &
      \subfigure[Coauthor-Physics]{
        \includegraphics[scale=0.33]{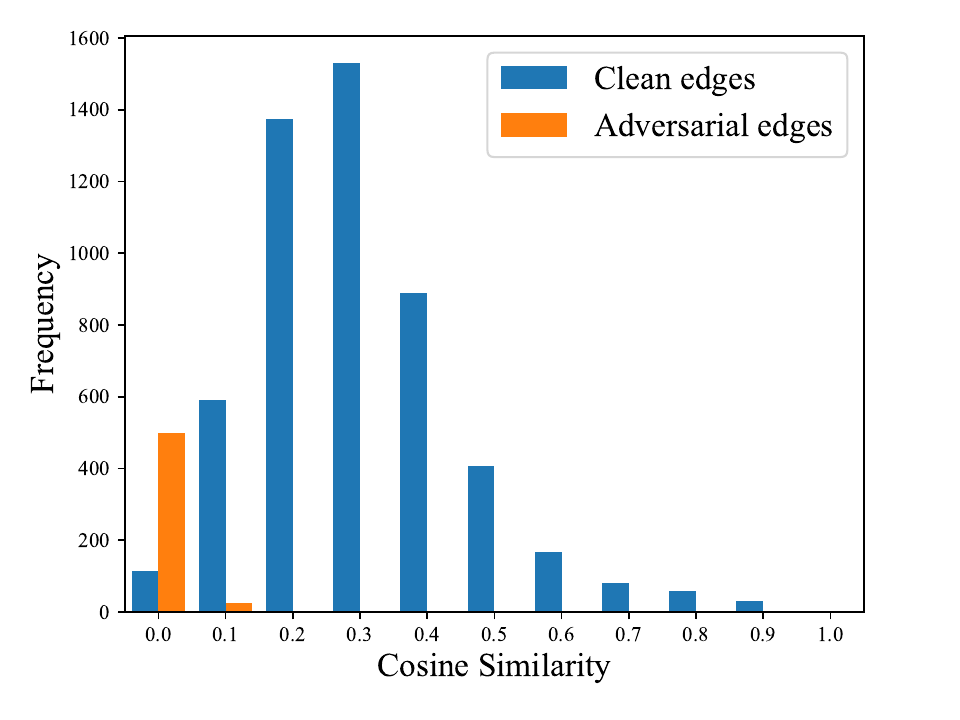}
        \label{fig:coauthor_physics}
      }  \\
    \end{tabular}
    \caption{Distribution of the cosine similarity of nodes connected by clean edges or adversarial ones.}
    \label{fig:append_cos_hist}
\end{figure*}

\figurename~\ref{fig:append_cos_hist} shows the distribution of the cosine similarity of neighboring nodes connected by clean edges or adversarial ones.
This distribution is based on the poisoned graphs created by maxmizing the loss of GCA.
We confirm that the poisoned graphs for ARIEL also have similar distribution in our preliminary experiment.
As shown in \figurename~\ref{fig:append_cos_hist}, adversarial edges on Cora, CiteSeer, Coauthor-CS, and Coauthor-Physics concentrate in the areas of lower values.
On the other hand, it is observed that adversarial edges on Amazon-Computers and Amazon-Photo are relatively widely distributed compared to other datasets.
According to these results, we consider that adversarial edges on the two Amazon datasets are more difficult to distinguish from clean edges even if the feature similarity is utilized.

\subsection{Hyperparameters of Existing Defense Methods for GNN} \label{appendix:GNN_baselines}
We utilize the codes shared by \cite{zhang2020gnnguard} to implement GNNGuard and GNNJaccard.
We use two-layers GCN in GNNGuard and GNNJaccard as with other GCL methods in our experiments.
Basically, we utilize the same hyperparameters used in the shared codes except for the dimension of the hidden layer.
We set the dimension of the hidden layer to 256 because the node classification accuracy is relatively low with a default value on some datasets.
Droprate, learning rate, and the number of epochs are 0.5, 0.01, and 200, respectively.
In all experiments, we conduct training models with Adam optimizer.
We repeat every experiment 10 times and report average results.

\subsection{Analysis of Pruned Edges in Poisoned Graph} \label{appendix:pruned_edges_analysis}
\begin{table*}[t]
  \caption{The ratio of adversarial edges to all the pruned edges (\%). The bold ratio is the highest one.}
  \label{tab:detection_rate}
  \centering
  \begin{tabular}{c|c|c|c|c|c|c}
    \toprule
    \textbf{Method} & \textbf{Cora} & \textbf{CiteSeer} & \textbf{Amazon-Computers} & \textbf{Amazon-Photo} & \textbf{Coauthor-CS} & \textbf{Coauthor-Physics} \\
    \midrule
    \prop\ (A) & 11.93 & \textbf{81.48} & 67.06 & 24.11 & 78.35 & \textbf{91.52} \\
    \prop\ (G) & \textbf{50.00} & 63.08 & 63.89 & 0.87 &  \textbf{80.10} & 31.93 \\
    \propnfs\ (A) & 29.17 &3.24 & \textbf{82.99} & \textbf{25.79} & 57.84 & 85.74 \\
    \propnfs\ (G) & 26.33 &18.46 &67.36 & 0.22 &53.56 &28.87 \\
  \bottomrule
\end{tabular}
\end{table*}
We analyze the reasons why the node classification accuracies on poisoned graphs are not completely improved to the level comparable to the ones on clean graphs even when our methods are applied to them.
We evaluate the ratio of successfully pruned adversarial edges to all the edges pruned by our methods.
\tablename~\ref{tab:detection_rate} shows the ratio of adversarial edges to all the pruned edges.
It is observed that the ratios are different depending on datasets.
According to \tablename~\ref{tab:poisoned_results} and \tablename~\ref{tab:detection_rate}, the higher the ratio is, the more the performance tends to be improved.
For example, \prop\ (A) can prune 81.48\% of adversarial edges on CiteSeer and achieves the best accuracy in \tablename~\ref{tab:poisoned_results}. 
These results mean that pruning more adversarial edges is important to restore the accuracy comparable to the accuracy on clean graphs. 
In order to further ameliorate the performance, other techniques may be needed, which is interesting and important future work.

\subsection{Interpretation of Improved Performance by Pruning Edges} \label{appendix:clean_improved_interpretation}
There are cases where the accuracies are improved when edges in clean graphs are pruned by \prop. 
In particular, \prop\ (A) and \prop\ (G) improve accuracies on both Coauthor-CS and Coauthor-Physics compared with ARIEL and GCA, respectively.
The reason for this could be that some noise edges included in clean graphs are removed.
Since our method prunes edges so that contrastive losses are minimized, such noise edges can be removed, which results in improving the performance of the node classification. 
However, on the other datasets, accuracies are degraded compared to existing GCL methods in almost all of the cases. 
This is because importance of edge information is different depending on the datasets.
According to experimental results in \cite{velivckovic2018deep}, when an unsupervised linear model is trained only with node features, node classification accuracies on Cora and CiteSeer are 47.9\% and 49.3\%, respectively.
When both edge information and node features are leveraged, the accuracy on Cora and CiteSeer are improved by approximately 35\%, and 20\%, respectively.
On the other hand, according to \cite{zhu2021graph}, node classification accuracies on Coauthor-CS and Coauthor-Physics are 90.3\% and 93.5\%, respectively even when only node features are utilized.
These results mean that edge information is less important on Coauthor datasets.
Thus, noise edges are relatively easily pruned on Coauthor datasets because useful node representation can be learned with node features, which results in improving the performance.
To the contrary, the more edge information is lost, the more difficult it is for an encoder to obtain useful node representation on other datasets. 
This is because the encoder has to learn complex information of nodes with node features and less edges.
As a result, since the quality of node embedding is easy to degrade, accuracy of node classification in downstream tasks may be degraded.

\subsection{Performance of Link Prediction} \label{appendix:link_prediction}
\begin{table*}[h!]
  \caption{Average AUC $\pm$ standard deviation (\%) of link prediction under CLGA.  The bold and underlined accuracies are the best and the second best ones, respectively.}
  \label{tab:auc_lp}
  \centering
  \begin{tabular}{c|c|c|c|c|c|c}
    \toprule
    \textbf{Method} & \textbf{Cora} & \textbf{CiteSeer} & \textbf{Amazon-Computers} & \textbf{Amazon-Photo} & \textbf{Coauthor-CS} & \textbf{Coauthor-Physics} \\
    \midrule
    \prop\ (A) & 98.02 $\pm$ 0.13 & 99.13 $\pm$ 0.09 & \textbf{94.93 $\pm$ 0.15} & 95.12 $\pm$ 0.13 & 99.08 $\pm$ 0.08 & 98.88 $\pm$ 0.13  \\
    \prop\ (G) & \textbf{98.49 $\pm$ 0.11} & 99.72 $\pm$ 0.05 & 94.25 $\pm$ 0.09 & 93.74 $\pm$ 0.13 & 99.53 $\pm$ 0.05 & 98.97 $\pm$ 0.06 \\
    \propnfs\ (A) & 98.04 $\pm$ 0.12 & 99.06 $\pm$ 0.08 & 94.85 $\pm$ 0.11 & 95.18 $\pm$ 0.05 & 99.14 $\pm$ 0.05 & 98.78 $\pm$ 0.06 \\ 
    \propnfs\ (G) & 97.94 $\pm$ 0.10 & 99.70 $\pm$ 0.02 & 94.50 $\pm$ 0.08 & 93.53 $\pm$ 0.10 & 99.38 $\pm$ 0.08 & 98.95 $\pm$ 0.11 \\
    \midrule
    \baseline\ (A) & 98.36 $\pm$ 0.10 & 99.34 $\pm$ 0.09 & \underline{94.86 $\pm$ 0.08} & \underline{95.56 $\pm$ 0.05} & 99.22 $\pm$ 0.03 & 99.15 $\pm$ 0.07 \\
    \baseline\ (G) & \underline{98.41 $\pm$ 0.13} & \underline{99.77 $\pm$ 0.03} & 94.37 $\pm$ 0.10 & 94.85 $\pm$ 0.09 & \textbf{99.55 $\pm$ 0.05} & \textbf{99.24 $\pm$ 0.07} \\ 
    
    \baselinenfs\ (A) & 98.36 $\pm$ 0.10 & 99.34 $\pm$ 0.09 & 94.77 $\pm$ 0.08 & \textbf{95.68 $\pm$ 0.04} & 99.22 $\pm$ 0.03 & 99.15 $\pm$ 0.07 \\
    \baselinenfs\ (G) & 98.41 $\pm$ 0.13 & \textbf{99.78 $\pm$ 0.03} & 94.45 $\pm$ 0.11 & 94.78 $\pm$ 0.11 & \textbf{99.55 $\pm$ 0.05} & \underline{99.23 $\pm$ 0.07} \\
    \midrule
    ARIEL & 97.85 $\pm$ 0.06 & 98.95 $\pm$ 0.08 & 94.46 $\pm$ 0.18 & 94.16 $\pm$ 0.03 & 98.97 $\pm$ 0.06 & 98.75 $\pm$ 0.16 \\ 
    GCA & 97.88 $\pm$ 0.10 & 99.65 $\pm$ 0.03 & 93.56 $\pm$ 0.09 & 92.23 $\pm$ 0.12	& 99.43 $\pm$ 0.04 & 98.90 $\pm$ 0.15 \\
  \bottomrule
\end{tabular}
\end{table*}
We evaluate link prediction performance of proposed variants on poisoned graphs.
We use a two-layer MLP as the projection head so as to project the embeddings into a new latent space following the setup in \cite{zhang2022unsupervised}.
For all the datasets, we split the edges into 70\%, 20\%, and 10\% for training, testing, and validating set, respectively. 
We report the area under curve (AUC) score. 
We repeat each experiment 5 times and report the average AUC and standard deviation.
\tablename~\ref{tab:auc_lp} shows the results of link prediction.
\prop\ (A) and \prop\ (G) achieve the best accuracies on Amazon-Computers and Cora, respectively.
On the other hand, baseline variants yield the best accuracies on CiteSeer, Amazon-Photo, Coauthor-CS, and Couthor-Physics.
From these results, baseline variants outperform \prop\ variants in terms of the link prediction task.
However, the difference is relatively marginal.
Furthermore, as mentioned in \Section~\ref{sec:eval_poisoned}, baselines are vulnerable to the adaptive attack described in \Section~\ref{sec:adaptive_attack}.
Therefore, we consider that \prop\ is more competent overall.

\subsection{Effects of Edge Addition on Poisoned Graph}  \label{appedix:amazon_study}
Although our method is dedicated to edge pruning, we evaluate the effects of adding edges on poisoned graphs of Amazon datasets.
We call the method that conducts both edge addition and deletion ``\propboth''.
\propboth\ mainly conducts edge addition in accordance with the conditions opposite to the ones for edge deletion.
The condition regarding gradients for selecting edges to add is represented as 
\begin{equation}
  \label{eq:add_condition}
 \SanitizedAdj[i,j] = 0 \land  \GradientSum[i, j] < 0.
\end{equation}
The condition regarding feature similarity for selecting edges to add is represented as
\begin{equation}
  \label{eq:fsim_condition_add}
 \CosSimilarity(\NodeFeature[i,:], \NodeFeature[j,:]) > \Threshold.
\end{equation}
On the other hand, the conditions for selecting edges to delete are Eq~(\ref{eq:del_condition}) and Eq~(\ref{eq:fsim_condition}) as with \prop.
We compare two types of \propboth\ in this experiment.
The first method simply modifies an edge that has the largest gradient out of edges satisfying the conditions.
We simply call this method \propboth.
The other method is \propbothbs\ that conducts edge modification on the basis of bernoulli sampling in order to avoid adding too much edges.
This is because deleting edges is more important from the perspective of defense.
To the contrary, deleting edges is less important from the perspective of the poisoning attack.
An adjacency matrix is basically sparse, which means that there are many zero values.
In other words, modifying values from 1 to 0 cannot effectively affect gradients of encoders because it is a trivial change in a clean graph.
As a result, edge addition tends to be conducted more frequently when poisoned graphs are created.
We tentatively leverage the ratio of existing edges (the value is 1) in a adjacency matrix as the probability of selecting addition.
This is because there the number of existing edges is less than non-existing edges in a sparse adjacency matrix, which is useful in controlling addition.
To be specific, the probability that edge addition is conducted is set to 0.0015 on Amazon-Computers.
On Amazon-Photo, we set the probability to 0.0022.
As with \prop, we evaluate four variants for each method depending on assumed GCL methods and using the feature similarity.
In other words, we run experiments on eight methods.
\tablename~\ref{tab:amazon_study} shows the results of average accuracy of node classification.
\propbothnfbs\ (A) achieves the best accuracies on both the datasets.
The second best accuracies are produced by \propbothbs\ (A).
The results demonstrate that controlling edge addition is more effective on Amazon datasets.
In particular, \propbothnfbs\ outperforms \propnfs\ (A), which attains the highest accuracies, 84.92\% and 88.58\% on Amazon-Computers and Amazon-Photo among \prop\ variants.
Thus, we consider that our method has potential for dealing with edges deleted by attacks although there is room for improvement.

\begin{table}[t]
  \caption{Accuracy of node classification on CLGA when both edge addition and deletion are allowed in \propboth. The bold and underlined accuracies are the best and the second best ones, respectively.}
  \label{tab:amazon_study}
  \centering
      \begin{tabular}{c|c|c}
        \toprule
        \textbf{Method} & \textbf{Amazon-Computers} & \textbf{Amazon-Photo}  \\
        \midrule
        \propboth\ (A) &  82.39 $\pm$ 0.91 & 87.60 $\pm$ 0.37  \\ 
        \propboth\ (G) &  81.02 $\pm $ 0.58 & 82.16 $\pm$ 0.71 \\ 
        \propbothnf\ (A) &  82.53 $\pm$ 0.71 & 87.87 $\pm$ 0.43  \\
        \propbothnf\ (G) &  78.39 $\pm$ 0.79  & 83.90 $\pm$ 0.47 \\ 
        \midrule
        \propbothbs\ (A) & \underline{83.16 $\pm$ 0.75} & \underline{88.77 $\pm$ 0.51} \\ 
        \propbothbs\ (G) &  81.16 $\pm $ 0.59 & 84.44 $\pm$ 0.44 \\ 
        \propbothnfbs\ (A) &  \textbf{85.20 $\pm$ 0.50} & \textbf{88.78 $\pm$ 0.30}  \\
        \propbothnfbs\ (G) &  82.82 $\pm$ 0.79  & 82.78 $\pm$ 0.52 \\ 
      \bottomrule
    \end{tabular}
\end{table}


\end{document}